\documentclass[nopreprintline,12pt]{elsarticle}
\usepackage{natbib}
\usepackage{amsmath,amssymb,amsfonts}
\usepackage{algorithmic}
\usepackage{graphicx}
\usepackage{textcomp}
\usepackage{tikz}
\usepackage{hyperref}
\usepackage{breqn}
\usepackage{multirow}
\usepackage{adjustbox}
\usepackage{subcaption}
\usepackage[T1]{fontenc}
\usepackage{float}
\def\ie{i.e.}

\def\etal{\emph{et al.}}

\newcommand{\paperName}{ReDMark} % DeepWater % ReDWater % DeepReDWater % DeepReDMark
\begin{document}

\begin{frontmatter}



\title{\paperName{}: Framework for Residual Diffusion Watermarking based on Deep Networks}


\author[lbl_iut]{Mahdi~Ahmadi}
\author[lbl_iut]{Alireza~Norouzi}
\author[lbl_michgan_compmed]{S.M.Reza~Soroushmehr\corref{cor1}}
\ead{ssoroush@med.umich.edu}
\author[lbl_iut]{Nader~Karimi}
\author[lbl_michgan_int,lbl_michgan_compmed]{Kayvan~Najarian}
\author[lbl_iut,lbl_michgan_int]{Shadrokh~Samavi}
\author[lbl_iut,lbl_quin]{Ali~Emami}
\address[lbl_iut]{Dept. of Electrical and Computer Engineering, Isfahan University of Technologyy, 84156-83111, Iran.}
\address[lbl_michgan_compmed]{Dept. of Computational Medicine and Bioinformatics, University of Michigan, Ann Arbor, 48109 U.S.A}
\address[lbl_michgan_int]{Michigan Center for Integrative Research in Critical Care, University of Michigan, Ann Arbor, 48109 U.S.A.}
\address[lbl_quin]{Dept. of Information Technology and Elect. Engineering, University of Queensland, QLD 4072, Australia}

\cortext[cor1]{Corresponding author.}

\begin{abstract}
Due to the rapid growth of machine learning tools and specifically deep networks in various computer vision and image processing areas, applications of Convolutional Neural Networks for watermarking have recently emerged. In this paper, we propose a deep end-to-end diffusion watermarking framework (\paperName{}) which can learn a new watermarking algorithm in any desired transform space. 
The framework is composed of two Fully Convolutional Neural Networks with residual structure which handle embedding and extraction operations. The whole deep network is trained end-to-end to conduct a blind secure watermarking. The proposed framework simulates various attacks as a differentiable network layer to facilitate end-to-end training. 
The watermark data is diffused in a relatively wide area of the image to enhance security and robustness of the algorithm. Comparative results versus recent state-of-the-art researches highlight the superiority of the proposed framework in terms of imperceptibility and robustness.
% Another important characteristic of the proposed framework, which leads to improved security and robustness, is its capability to diffuse watermark data among a relatively wide area of the image. 
% The framework is customizable for the level of robustness vs. imperceptibility. It is also adjustable for the trade-off between capacity and robustness. 
\end{abstract}

\begin{keyword}
Blind watermarking \sep data diffusion \sep deep convolution networks \sep 
\sep CNN \sep FCN \sep transparency. 
\end{keyword}

\end{frontmatter}


\section{Introduction}
\label{sec:introduction}
Digital watermarking was originally introduced in 1979 for anti-counterfeit purposes \cite{p1} to distinguish between the original and counterfeit documents. Since then it has been applied for identification of image ownership and protection of intellectual property by hiding data such as logos and proprietary information in images, videos and audios \cite{p2}. Another application is the patient identification and medical procedure matching by hiding patients' personal information in their medical images \cite{p3}.  Other applications have been proposed for watermarking such as broadcast monitoring \cite{p4}, copy control \cite{p5}, device control \cite{p6} and legacy enhancement \cite{p7}. 
The most well-known challenge in watermarking is that watermarked image which contains hidden data is vulnerable to image processing algorithms for enhancement, transformations like image compression and format conversion and undesired artifacts such as transmission noises. 
Furthermore, watermarked images are prone to intentional attacks which strive to change or corrupt the hidden watermark data.
Despite the extensive amount of research to battle these problems, robustness, and imperceptibility are still the two key challenges in watermarking algorithms. In another word, one major concern in modern watermarking is to preserve the hidden data as safe as possible in the presence of attacks (robustness), while introducing subtle and undetectable changes during the watermarking process, so that the watermarked image would be indistinguishable from the original image (imperceptibility). Another problem which has attracted a great deal of research over the last two decades is the blindness of watermarking algorithms, which increases their complexity and may negatively affect their robustness and imperceptibility. However, the blind watermarking methods \cite{p9,p10} are practically preferable to informed/non-blind methods \cite{p11,p12}, since the informed methods require various side-information about watermark or cover image or embedding parameters for extraction.

Nowadays the application of machine learning tools in watermarking is growing very rapidly, because of their effective solutions to embedding and extraction processes \cite{p14,p15,p16,p17,p18,Heidari}.  Nevertheless, most of them generally utilize machine learning tools such as Support Vector Machine (SVM) \cite{SVM}, Support Vector Regression (SVR) \cite{SVR}, Radial Basic Function Neural Network (RBFNN) \cite{RBFNN}, and K Nearest Neighbor (KNN) \cite{KNN} for specific parts of watermarking procedure such as parameter optimization \cite{p14,p15}, prediction of transform domain coefficients \cite{p16,p17,p18} and attack estimation \cite{Heidari}. Among all the machine learning tools, deep networks and Convolutional Neural Nets (CNN), have gained the most widespread attention in a large variety of computer vision applications such as pattern recognition \cite{p20}, image classification \cite{p21} and object detection \cite{p22}. Very recently a few works have emerged about the application of deep networks in watermarking \cite{p24,p25,Hidden}. Kandi \etal \cite{p24} proposed CNN based auto-encoder structures to hide watermark data in their feature maps. However, their proposed watermarking method is non-blind, and a predefined embedding algorithm is applied for embedding in auto-encoder feature maps. In \cite{p25}, an end-to-end watermarking network is introduced. However, the watermark data is embedded in single blocks of the image, which leads to a uniform local embedding similar to traditional methods.

Parallel to Li \etal \cite{Hidden}, who have presented a unified system for watermarking and steganography based on CNNs and Generative Adversarial Networks \cite{GAN}, in this paper we introduce an end-to-end blind watermarking framework (\paperName{}) using Fully Convolutional Neural Networks (FCN), which is capable of learning a new watermarking algorithm and surpass state-of-the-art (including \cite{Hidden}) in terms of robustness and imperceptibility. The proposed system consists of two FCNs for embedding and extraction, along with a differentiable attack layer which simulates well-known attacks. Introducing differentiable attack layer as part of the network makes an end-to-end training scheme feasible.  It also leads to robust watermarking due to training the network in the presence of attacks. To this end, we suggest a differentiable approximate model for JPEG attack with the adjustable quality factor.

Dominant watermarking approaches use fixed methods such as swapping coefficients in a transform domain. Nevertheless, \paperName{} is capable of learning several embedding patterns/masks in different transform domains and in presence of various attacks. Consequently, the network explores suitable solutions customized for the suggested transform domain and required attacks. On the other hand, only the constructed network can embed and extract the watermark data based on the discovered watermarking patterns. Hence, the proposed system introduces a secure method for hiding the watermark data so that the recognition or replacement of the secret data along the communication channel is not easy.

Another important characteristic of the suggested system, which leads to improved security and robustness, is its capability to diffuse watermark data among a relatively wide area of the image. In other words, the network explores diffusion watermarking masks to share watermarking data among several image blocks, rather than simply swapping in a single block. Thus, the watermarked image demonstrates impressive robustness against several heavy attacks. Even if a meaningful part of image is corrupted or removed, the extraction network is still able to extract the hidden watermark.

The last elegant feature of \paperName{} is a strength-factor for controlling the strength of the watermark patterns within the image, thanks to the novel structure of the embedding network which is inspired by ResNet \cite{Resnet}. This valuable feature enables us to control the trade-off between the robustness and imperceptibility depending on situation and application requirements.

To make a long story short, our major contributions in the introduced deep watermarking network include: 1. Proposing a residual watermarking framework with specialized robustness against several specific attacks. 2. Introducing a strength factor tuner for controlling the trade-off among robustness and imperceptibility. 3. Introducing a new differentiable approximation of JPEG attack with any quality factor. 4. Proposing a novel diffusion watermarking framework built on circular convolutional layers which leads to exceptional robustness against various attacks.

The remainder of the paper is arranged as follows: In section \ref{sec:Background} we review related works in the literature. Technical details of the proposed framework are discussed in section \ref{sec:Technical}. The experimental results are presented in section \ref{sec:Experiments}. Finally, we conclude the paper with a short discussion in section \ref{sec:Conclusion}.
\section{Related Work} 
\label{sec:Background}

Since the advent of digital watermarking, an enormous amount of research has been invested in developing new watermarking schemes with improved capacity, robustness, imperceptibility, fidelity and security. Early methods embedded the watermark data in the spatial domain by directly manipulating image pixels to represent watermark data. Embedding watermark bits in LBS of image pixels is an example of such methods \cite{p28}. For improving robustness, the mainstream in literature is to embed the watermark data in a transform domain by manipulating specific transform coefficients. Some popular transform domains suggested for watermarking are DCT \cite{p29}, Wavelet \cite{p30}, Hadamard \cite{p31}, Contourlet \cite{p32} or a mixture of transforms \cite{Fazlali}. Sadreazami \etal \cite{p34} proposed a multiplicative embedding method in Contourlet domain, which requires statistical analysis for data extraction. They simulate the behavior of watermarked image as normalized inverse Gaussian distribution and propose a maximum likelihood detector for data extraction. Makbol \etal \cite{p35} utilize integer Wavelet transform and singular value decomposition for watermarking. Another research \cite{p36} suggested a reversible transform based on an over-complete dictionary for watermarking. The authors in \cite{p37} introduced quaternion Hadamard transform domain for watermarking in color images by using Schur decomposition. Liu \etal \cite{p38} proposed a transform domain called fractional Krawtchouk transform for watermarking. They use Dither modulation method \cite{p39} for embedding in this domain.

Following increased applications of machine learning tools for various tasks, some researchers started to apply these tools for different parts of embedding and extraction in the watermarking process. For example, Heidari \etal \cite{Heidari} presented a framework for blind image watermarking by the redundant embedding of watermark data in multiple zones of the DCT spectrum. For extracting watermark data from the attacked image, they apply SVM to recognize the least distorted zone of the spectrum. In \cite{p14}, K-NN regression method is utilized for estimating the optimum value of embedding strength parameter in DCT domain for improving robustness and imperceptibility. Zhi-Ming \etal \cite{p15}, suggest to use RBFNN for optimization of embedding strength in blocks of the cover image. The embedding strength for each block is determined separately based on the block features obtained by RBFNN. Authors in \cite{p16} use Wavelet domain for watermarking. They predict some coefficients by SVR and use an embedding rule based on the predicted values versus the real ones. Likewise, in \cite{p17}, Lagrangian support vector regression is utilized for prediction of coefficients and embedding process in lifting Wavelet transform space \cite{p40}. Furthermore, a scaling factor is assigned for embedding strength in each block which is estimated by genetic algorithm. A similar embedding method is utilized by \cite{p18}, which propose Extreme Learning Machine (ELM) for prediction in the DWT domain. In spite of vast research proposals for the application of machine learning tools in watermarking, none of the above-mentioned methods propose a unified watermarking framework based on machine learning approaches. In other words, all of them apply machine learning tools on specific parts of the watermarking process.

Among all machine learning tools, CNNs have gained extreme popularity in the last decade. However, their assistive application for the watermarking process is more recent. For example, Kandi \etal \cite{p24} applied two auto-encoder CNN structures for feature extraction to be separately used for positive and negative embedding. The same auto-encoder networks are used at the receiver side to obtain the feature maps and extract the watermark data. In \cite{p25}, two CNN networks are trained for embedding and extraction in the presence of attacks. However, the watermarking networks are designed for hiding a one-bit watermark in a single block. The most relevant research to ours is the end-to-end trainable framework, HiDDeN \cite{Hidden} which is proposed for data hiding in color images based on CNNs and GANs and may be applied to watermarking and steganography. A noise layer is proposed in the network for simulating attacks during the end-to-end training phase.
 
\section{Proposed Watermarking Framework}
\label{sec:Technical}
\begin{figure*} [!tb] 
	\begin{minipage}{1.9\textwidth}	
	  \begin{minipage}{0.5\textwidth} 
	  			\centering
				\includegraphics[width=1.0\textwidth, height=5cm,keepaspectratio]{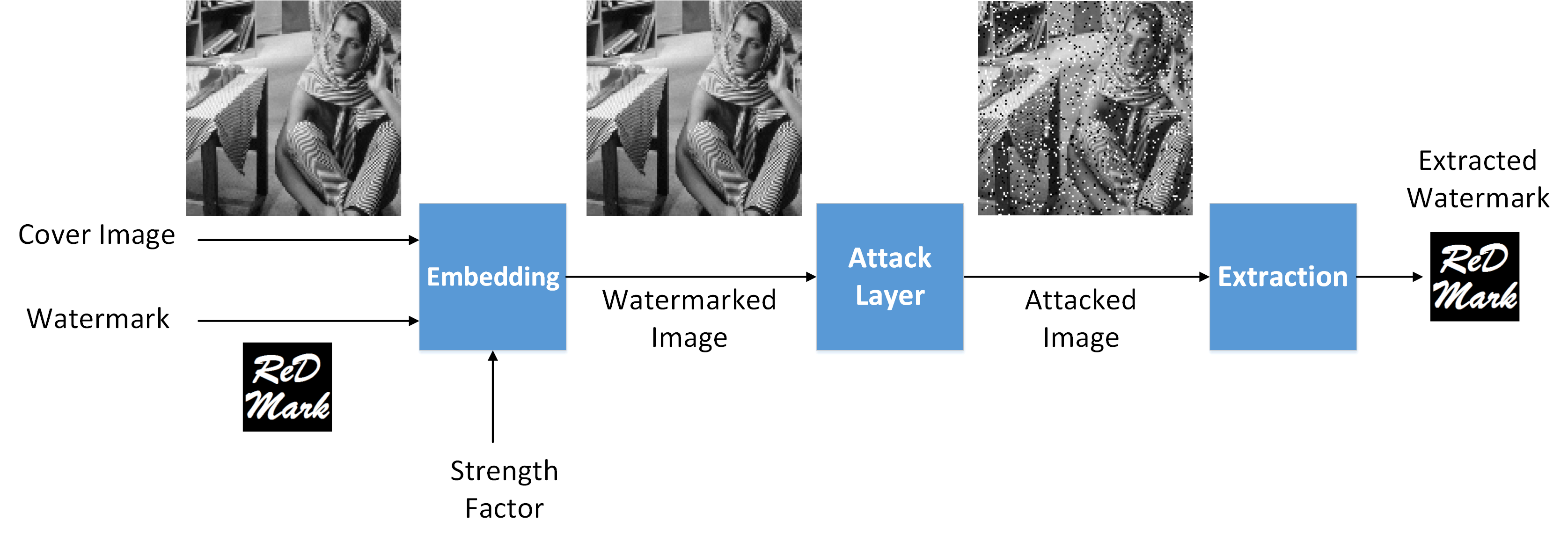} 
	\end{minipage}
\end{minipage}
  \caption
    {\small	Block diagram of the proposed watermarking framework.}
  \label{fig:overall_arch}
\end{figure*}
In this paper, we propose an adaptive diffusion watermarking framework (\paperName{}) composed of two Fully Convolutional Networks with residual connections. The proposed networks are trained end-to-end to conduct a blind secure watermarking for grayscale images in the desired transform space. The framework is customizable for the level of robustness vs. imperceptibility. It is also adjustable for the trade-off between capacity and robustness. The adaptive and flexible nature of the framework makes it easy to choose any linear transform domain for embedding the secret watermark or to train the watermarking network for higher resistance to specific attacks. Employing a differentiable attack module as part of the network facilitates end-to-end training and governs robust watermarking against various attacks. We elaborate on the technical details of the system modules and their functionalities in section \ref{sec:structure} and discuss the end-to-end training strategies in section \ref{sec:strategy}.

\subsection{Network Structure}
\label{sec:structure}
Fig. \ref{fig:overall_arch} illustrates a block diagram of the system composed of three main modules: CNN for embedding the watermark, differentiable attack layer for simulating popular attacks, and CNN for extraction of the hidden watermark.
\vspace{3pt}
\subsubsection{Embedding Module}
\label{sec:Embedding}
\begin{figure*} []
	\begin{minipage}{1.9\textwidth}	
	  \begin{minipage}{0.5\textwidth} 
	  			\centering
				\includegraphics[width=1.0\textwidth, height=6cm,keepaspectratio]{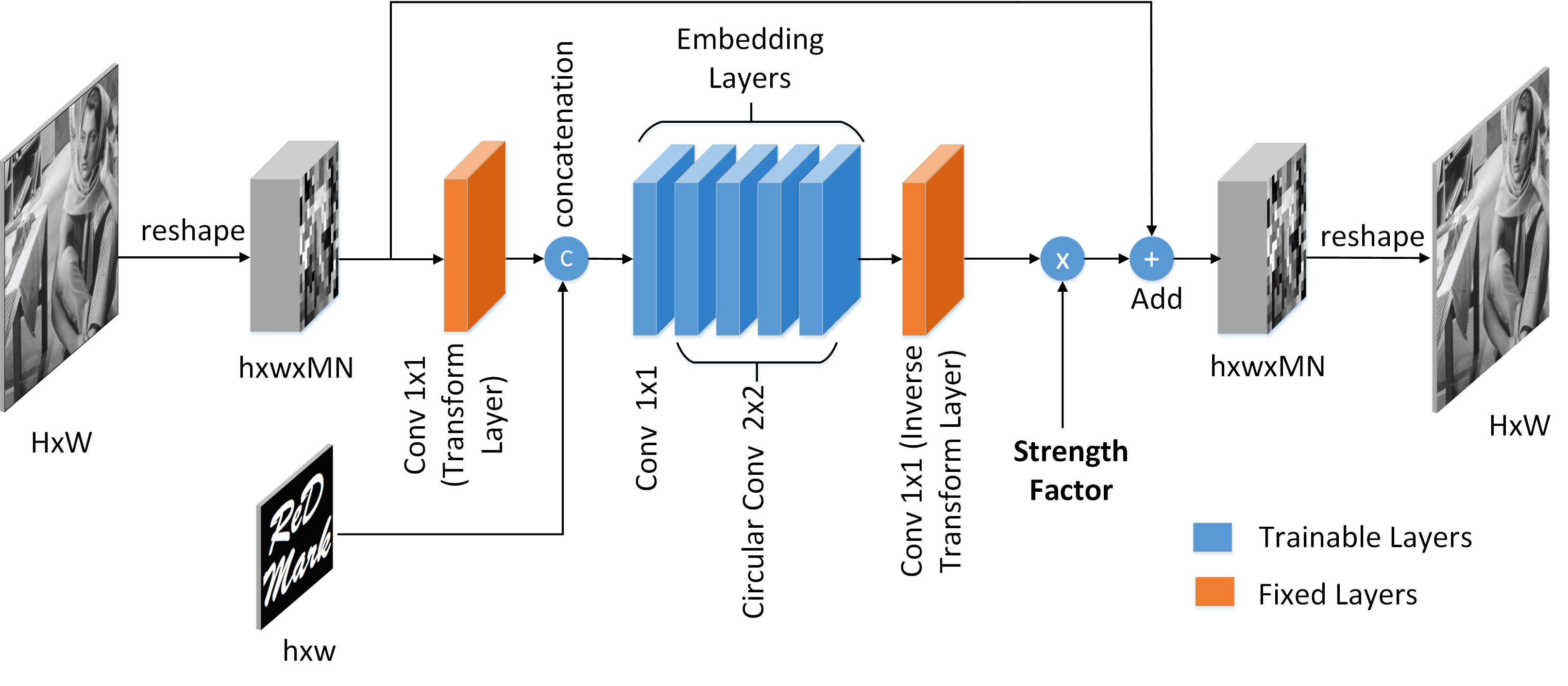} 
	\end{minipage}
\end{minipage}
  \caption
    {\small Embedding network: Trainable layers and fixed transform layers are shown in different colors (Fixed layers may be initialized and released for training as well).}
  \label{fig:embedding}
\end{figure*}

As shown in Fig. \ref{fig:embedding}, the embedding network structure is composed of two transform layers, and five convolutional layers. For a $H\times W$ grayscale cover image and $h\times w$ binary watermark data, the embedding network embeds $h\times w$ bits of watermark in a bigger cover image with $H\times W$ pixels ($Capacity=\frac{h\times w}{H\times W}(bpp)$). The embedding layers, compute the watermarking mask/pattern within the transform domain. Then the residual mask is calculated in the spatial domain by the inverse transform to be added to the original image with a Strength Factor ($\alpha$) weight. Further technical details of the pipeline are discussed as follows.

\paragraph*{Space to Depth (Reshape)}
Let's assume the cover image can be properly divided into blocks of size $M\times N$ and each block will be hosting at least one watermark bit.  Without loss of generality, we assume that the cover image has $h\times w$ blocks required for the watermark bit length.  Hence, we reshape the cover image into a tensor of size $MN\times h \times w$. Each column of the generated tensor is the vectorized form of $M\times N$ image blocks.

\paragraph*{Transform Layer}
This layer implements a reversible linear transform to change the representation basis of the image from the spatial domain to a new space such as a frequency domain. Although watermarking in spatial domain provides higher capacity with lower complexity, it is shown that watermarking in a transform domain is more secure and robust against intentional or random attacks and image processing techniques \cite{p41}.  To perform the watermarking process in a new domain, we need two transform and inverse transform layers for input and output interface of the embedding network. The suggested transform layers can be fixed to any standard transform such as DCT, wavelet or Hadamard. However, it is also possible to pre-assign an arbitrary transformation to the layer and let the network fine-tune the transform layer throughout the training process within a specific training strategy. Considering the proposed rearrangement of the cover image in the previous paragraph, the transformation layer is simplified into a $1\times 1$ convolution layer, as we apply the transform on each block independently. Depending on the new space dimension, we need $n_T$ convolution masks in accordance to $n_T$ transformation basis. Hence, every $M\times N$ block is reshaped to $1\times 1\times MN$ tensor and convolved with $1\times 1$ transformation masks. Output of each of $n_T$ masks is calculated by:
\begin{eqnarray}
 \label{prelema} 
\mathbf{f_T}=\left[\mathbf{f_T}(\theta)\right]_{n_T\times 1}=\left[   \sum_{k=1}^{MN}{\mathbf{f}(k)\mathbf{D}(\theta,k)}   \right]_{n_T\times 1}=\nonumber\\
\left[\mathbf{d}_\theta\right]_{n_T\times MN}\left[\mathbf{f}\right]_{MN\times 1}=\mathbf{D}_{n_T\times MN} \mathbf{f}_{MN\times 1}
\end{eqnarray}
where $\mathbf{f}$ is the $1\times 1 \times MN$ block tensor reshaped as a column vector, $\mathbf{d}_\theta$ is the vectorized $1\times 1$ convolution mask representing the weights of one neuron and $\mathbf{f_T}(\theta)$ is the output of $\theta^{th}$ filter mask. $\mathbf{D}=\left[ \mathbf{d}_\theta \right]$ represents the $n_T\times MN$ transform matrix, where its rows contain the corresponding neuron weights and $\mathbf{f_T}$ demonstrates the transformed feature space, \ie, outputs of all neurons. There are no bias values for neurons.
Special cases of Equation \eqref{prelema} for DCT and Hadamard transforms are discussed in the appendix. Similar calculations are required for any other linear transform space.

There is no obligation for using a fixed known transform in the transformation layers. To elaborate, the transform layers of the proposed framework may be initialized to any desired transform or even random values and released for training during end-to end training of the whole network. We only need to constrain the two input and output transform layers of the two embedding and extraction networks to be equivalent inverse transforms. In this way we expect the network to seek new transform domains for watermarking. However, in this work we use a fixed transformation layer for proof of the framework concept.

\paragraph*{Embedding layers}
The output of the transformation layer is concatenated with the watermark image, shaping the input tensor of size $(MN+1)\times h\times w$ for the embedding network. This network is composed of circular and normal convolutional layers, which are responsible for embedding the watermark patterns into the transformed image blocks. As shown in Fig. \ref{fig:embedding}, some layers of the embedding network perform $2\times 2$ circular convolution, which lead to expanding the receptive field of neurons in the final layers. Hence, this innovative structure empowers diffusion watermarking, so that the watermark data is shared and distributed among adjacent blocks. Furthermore, the embedded watermark added to each block of the cover image is a superposition of symbols in the wide receptive field, \ie, the own block and its neighbors. This brilliant property improves security and robustness of the proposed framework against several attacks.

\paragraph*{Skip Connection and Strength Factor}
As displayed in Fig. \ref{fig:embedding}, the watermarked image is produced by summing the output of the embedding network with the original cover image. This structure guides the network to produce the residual watermark data. This helps the embedding network to learn the additive watermarking symbols more efficiently, \ie, the network weights converge faster during the training stage. On the other hand, the proposed network structure empowers the framework to incorporate a Strength Factor  to adjust the strength of generated symbols before summation with the cover image. Addition of this elegant tuning volume to a trained network enables the system to amplify or attenuate the generated symbol in the watermarked image and control the level of robustness vs. imperceptibility (PSNR/SSIM) based on our requirements. During the network training, strength factor is fixed to one.
\vspace{3pt}
\subsubsection{Attack Layer}
\label{sec:Attack}
In this part, we elaborate on the structure of the attack layer as shown in Fig. \ref{fig:overall_arch}.  The proposed framework simulates various attacks as a differentiable network layer to facilitate end-to-end training.  Furthermore, keeping the attacks in the training loop guides the network to learn more robust watermarking patterns to resist real-world attacks in communication channels. A network trained in the presence of an attack will produce watermarked images that are more robust to that specific attack.

Interestingly, learning robust watermarking for specific attacks may lead to robustness against some other attacks, due to similarities in their natures. In this work, we train separate networks with well-known attacks and analyze their robustness against many other attacks. We also train a network with multiple simultaneous attacks which is shown to resist a wider range of attacks simultaneously. Various attacks simulated for network training in this work are briefly explained below, and details of the training are discussed in section \ref{sec:Experiments}.

\paragraph*{Noise Attack}
This is a random white noise added to the watermarked image in every iteration of training. Hence, the back-propagated loss signal passes through the additive noise which is fixed for every iteration. We may apply various noise types such as uniform noise and Gaussian noise. Likewise, salt \& pepper noise follows the same logic, where either of values 0 or 255 are assigned to random image pixels with a specific probability.

\paragraph*{Random cropping attack}
This improves watermarking performance in the presence of cropping attack since the network learns to redundantly embed watermark data in different regions. Random cropping is implemented by suppressing or turning off neurons in random block regions. The process is similar to dropout layers introduced in \cite{Dropout}. 

\paragraph*{Smoothing Attack}
We use a normalized unit mask (all-ones matrix) for smoothing  windows, as shown in Equation \eqref{smoothing}.
\begin{equation} \label{smoothing}
\mathbf{H}_{mask}=\frac{1}{a^2} 
\begin{bmatrix}
1 & \dots  & 1\\
\vdots  &\ddots  & \vdots\\
1 & \dots & 1 \\
\end{bmatrix}_{a\times a}
\end{equation}
Thus, the attack layer, in this case, is a simple convolutional neuron with constant weights. We may use any convolution mask for various FIR (Finite Impulse Response) filters, to simulate any band filtering attack. For example, the layer mask can be set to Gaussian filter or a sharpening filter.

\begin{figure}[]
  \centering
    \includegraphics[width=1.0\columnwidth,height=5cm,keepaspectratio]{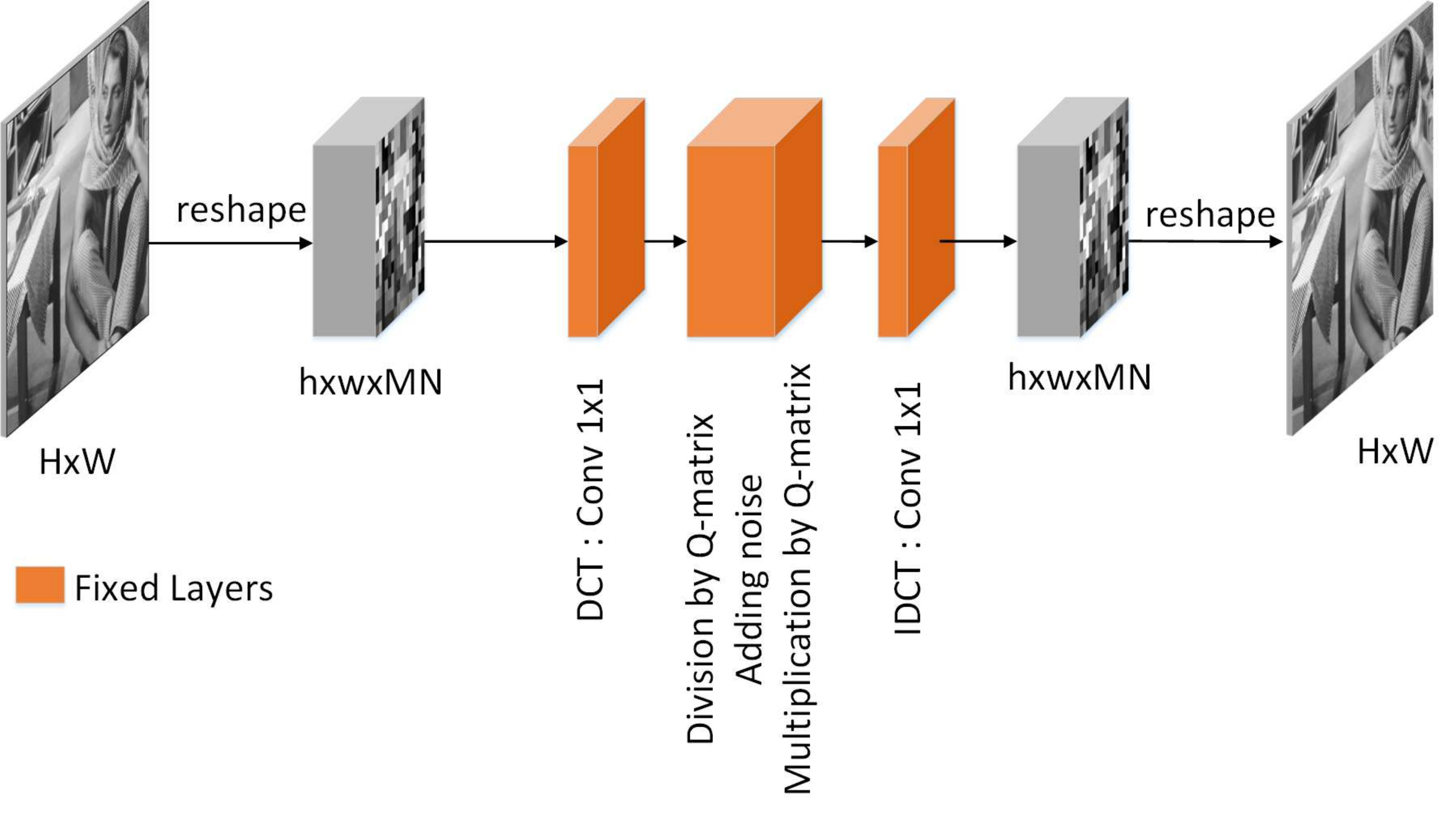}
    \caption{\small Differentiable approximate model for JPEG attack layer}
    \label{fig:diffJpeg}
\end{figure}

\paragraph*{JPEG Attack}
All the attacks discussed so far, are inherently differentiable and can be directly implemented into the network layer. However, JPEG coding involves some non-differentiable operations. Thus, we need a differentiable approximation of the process to simulate the JPEG attack in a network layer. The JPEG compression includes the following steps: transferring image blocks to DCT domain, dividing by a quantization matrix determined by a quality factor and rounding the results to integer values. Similarly, JPEG decoding involves inverse operations as follows: Multiplying the coefficients by the same quantization matrix, then transforming to the spatial domain. A complete simulation of JPEG attack, as shown in Fig. \ref{fig:diffJpeg}, consists of all the stages mentioned above: DCT transform, division by quantization matrix, rounding, multiplication by quantization matrix, inverse DCT transform (see Transform Layer in section \ref{sec:Embedding}. Among all the mentioned steps, rounding operation is non-differentiable and needs to be approximated with a differentiable operation, to facilitate back propagation of the training gradients.  We simulate rounding operation by a uniform noise in the range [-0.5, o.5]. Mathematical rounding is practically a subgroup of the suggested approximation, in other words, we are simulating a larger family of distortions than normal JPEG. Equation \eqref{Jpegeq} demonstrates the proposed rounding simulation method:
\begin{equation} \label{Jpegeq}
\mathrm{Iw}_{DCT}^*=\left(\frac{\mathrm{Iw}_{DCT}}{Q}+\sigma  \right)\times Q=\mathrm{Iw}_{DCT}+\sigma Q
\end{equation}
where $\mathrm{Iw}_{DCT}$ is the watermarked image in the DCT domain, $\mathrm{Iw}_{DCT}^*$ is its approximated quantized version, $Q$ is the quantization matrix of a required quality factor and $\sigma$ is the uniform noise [-0.5, o.5]. The equation implies that the rounding effect appears more strongly for larger elements of the quantization matrix, which are towards higher frequencies. This guides the network to gradually reduce the embedding strength in higher frequencies of the transform domain.

\begin{figure}[!h]
	\centering
	\includegraphics[width=0.7\columnwidth]{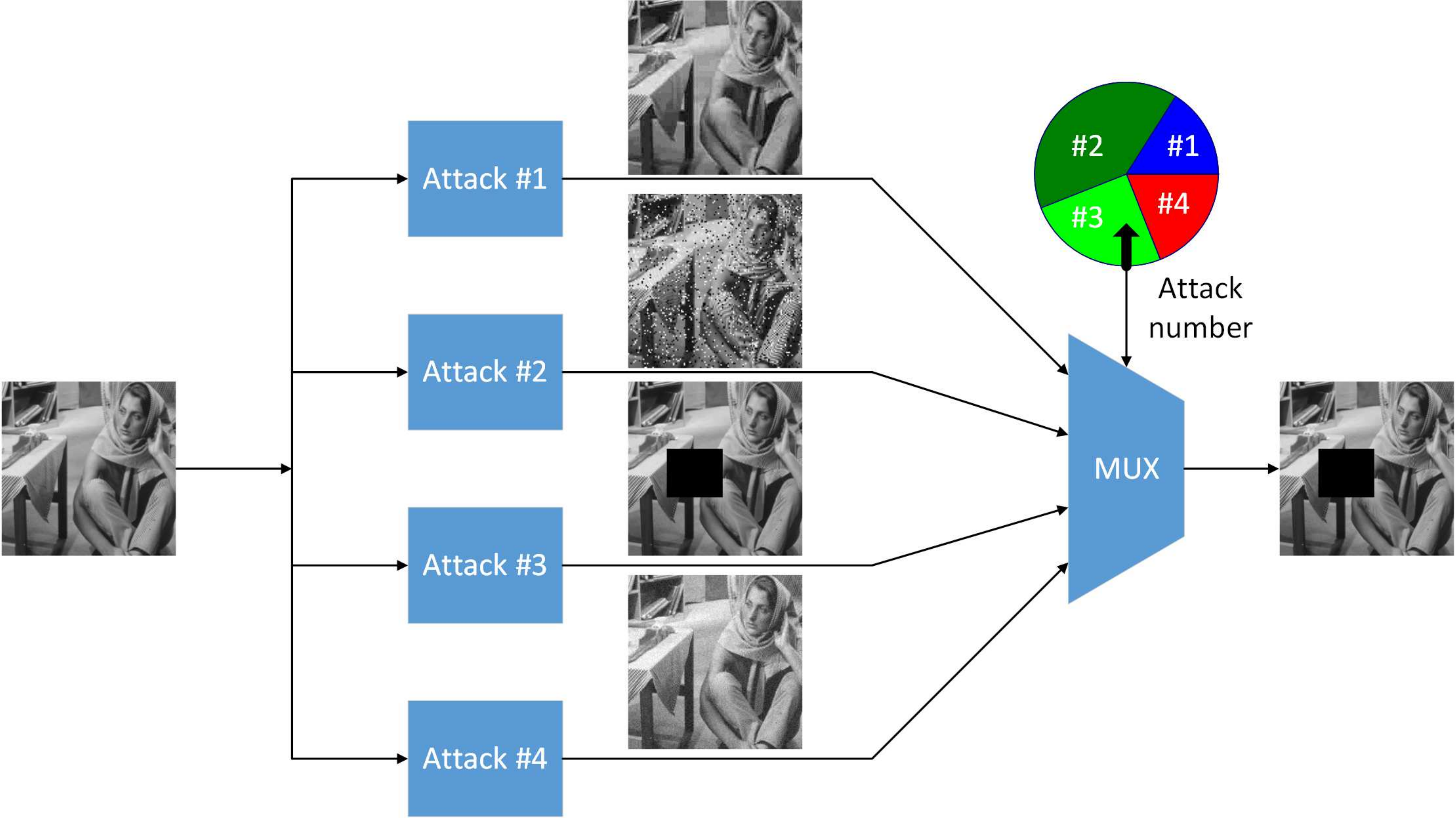}
	\caption{\small Multi-attack layer: In every iteration, only one type of attack is applied on the watermarked image}
	\label{fig:multi_switch}
\end{figure}

\paragraph*{Mixture Of Attacks}
Multiple attacks may be combined in the attack layer to train a robust watermarking network for the mixture of chosen attacks. The training procedure, in this case, is slightly different, as in each iteration of training the network randomly selects one of the attacks with a given probability. Hence, the back propagated gradients are passed through the selected attack layer. The switching mechanism of the multi-attack layer is illustrated in Fig. \ref{fig:multi_switch}. As shown in the figure, we model the random selection of attacks as a roulette wheel which assigns a probability to each attack. Thus, in each of the training iterations, only one type of attack is allowed to pass through the multiplexer to affect the training loss.

%\vspace{3pt}

\subsubsection{Extraction Module}
\begin{figure}[]
  \centering
    \includegraphics[width=1.0\columnwidth,height=4cm,keepaspectratio]{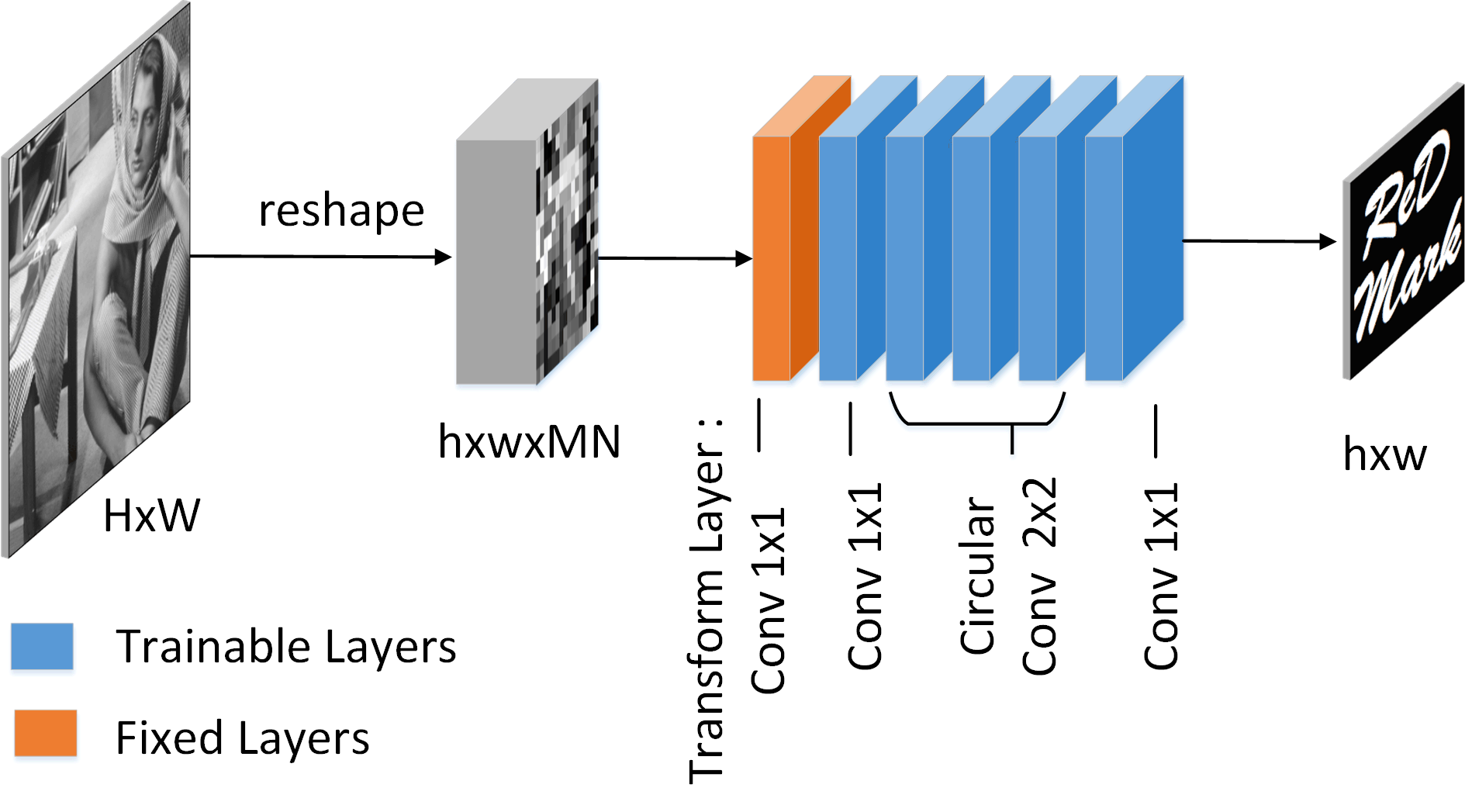}
    \caption{\small The extraction network}
    \label{fig:extraction_net}
\end{figure}

As illustrated in Fig. \ref{fig:extraction_net}, the extraction module is structurally simpler than the embedding network. This module is supposed to extract watermark data from the input image. Since the watermark data is embedded in the transform domain, the extraction module incorporates a copy of the transform layer utilized for the embedding module, to represent the watermarked image in the same basis. Other network layers learn to extract the watermark data in the transform domain.

\subsection{Network Training and Evaluation Metrics}
\label{sec:strategy}
As can be seen in Fig. \ref{fig:overall_arch}, the trainable network modules (embedding and extraction units) are trained together within an end-to-end setup including the trainable and non-trainable network layers. The main training objectives are to establish embedding and extraction networks which can generate safe, high-quality watermark images and robustly recover the hidden data from the watermarked images. In other words, each network has an independent objective function. The embedding network is supposed to generate a watermarked image with maximum quality and minimum distortion compared to the original image. On the other end, the extraction network is responsible for maximizing the extraction rate of the hidden watermark or equivalently minimizing Bit Error Rate (BER), as defined by Equation \eqref{BER}.
\begin{equation} \label{BER}
\mathrm{BER}(W_t,W'_t)=\frac{\sum_{l=1}^{L_{\mathrm{w}}}{\mathrm{XOR} (W_t(l),W'_t(l)) }}{L_{\mathrm{w}}}
\end{equation}
where $W_t$ is the original binary watermark and $W'_t$ is the extracted watermark. $L_{\mathrm{w}}$ represents the watermark string length. This $L_{\mathrm{w}}$ bits of watermark may be embedded redundantly in the cover image. Further discussion about redundant embedding is in section \ref{sec:expsetup}. We utilize two metrics for evaluating the image quality in training and test stages. The Structural Similarity Index (SSIM) is a perceptual metric employed as a training loss function to quantify degradation of image quality caused by the watermarking process and transmission. SSIM estimates the structural variation of the two images through the Equation \eqref{SSIM}:
\begin{equation} \label{SSIM}
L_1=\mathrm{SSIM}(\mathrm{I},\mathrm{I_w})=\frac{(2\mu_{\mathrm{I}} \mu_{\mathrm{I_w}}+c_1) (2\sigma_{\mathrm{I},\mathrm{I_w}}+c_2)}{(\mu_{\mathrm{I}}^2 \mu_{\mathrm{I_w}}^2+c_1)(\sigma_{\mathrm{I}}^2 \sigma_{\mathrm{I_w}}^2+c_2)}
\end{equation}

where $\mathrm{I}$ is the cover image, $\mathrm{I_w}$ is the watermarked image, $\mu_{\mathrm{I}}$  and $\mu_{\mathrm{I_w}}$ are the mean values of $\mathrm{I}$ and $\mathrm{I_w}$ respectively, $\sigma_{\mathrm{I}}$  and $\sigma_{\mathrm{I_w}}$ represent their variances and $\sigma_{\mathrm{I},\mathrm{I_w}}$ is the covariance of $(\mathrm{I},\mathrm{I_w})$. In this equation $c_1$ and $c_2$ are two constants of the metric which are set to $10^{-4}$ and $9\times 10^{-4}$ for our experiments.

For comparing the watermarking quality of the \paperName{} (imperceptibility of the produced watermarked images) against state-of-the-art competitors, we exploit the well-known PSNR metric (Peak Signal-to-Noise Ratio), as shown by Equation \eqref{PSNR}:
\begin{equation} \label{PSNR}
\mathrm{PSNR}(\mathrm{I},\mathrm{I_w})=10\log{\left(  \frac{W \times H \times MAX^2}{\sum_s{\sum_t{\left\vert  \mathrm{I}(s,t)-\mathrm{I_w}(s,t) \right\vert^2}}}  \right)}
\end{equation}
where $W$ and $H$ are the image dimensions, and $MAX$ stands for the maximum value of image pixels (for grayscale images, $MAX=255$).

Based on the above discussion, for the end-to-end training of the multi-objective network, we employ a weighted combination of the embedding and extraction loss functions, as shown in Equation \eqref{losses}. 
\begin{equation} \label{losses}
L=\gamma L_1 + (1-\gamma) L_2
\end{equation}
where $\gamma$ is the ratio of losses and $L_1,L_2$ are the loss functions of the embedding and extraction networks. $L_1$ simulates imperceptibility or quality of the watermark image, while $L_2$ represents the watermark extraction rate and robustness. We use SSIM metric for $L_1$ and binary cross entropy for $L_2$:
\begin{align} 
\label{crossentropy}
L_2=-\sum_{\substack{all\\pixels} }{\sum_{c=1}^2{y_c\log (p_c)}}=-\sum_{\substack{all\\pixels} }{y_1\log(p_1)+(1-y_1)log(1-p_1)}
\end{align}
where $p_i$ values are the outputs of the extraction network representing the probability of watermark bits. Watermark data is generated by thresholding these values.

Since there is a trade-off between the two loss-functions (imperceptibility vs. robustness), the process of network training is a multi-objective optimization. The flow of gradients from the last layers of the extraction network to the first layers of the embedding network implies that only $L_2$ gradients back-propagate through the extraction network. However, for the embedding layers in the head of the pipeline, gradients of the combinatorial loss $L$ back-propagate to train the embedding network weights.
 
\section{Experimental Results}
\label{sec:Experiments}
The proposed framework is implemented by Tensor-flow \cite{Tensorflow} and executed on NVIDIA GeForce\textsuperscript{\textregistered} GTX 1080 Ti. We use CIFAR10 \cite{CIFAR} and Pascal VOC \cite{Pascal} datasets for training the network. The dataset of 49 standard test images from the University of Granada \cite{dataset} is used for most of the numerical analysis and comparative evaluations. We demonstrate the superiority of \paperName{} by comparing with two state-of-the-art watermarking systems \cite{Hidden}, \cite{p36}. For a fair comparison against the concurrent work of Zhu \etal \cite{Hidden}, we evaluate our system on COCO dataset \cite{COCO}.

To demonstrate the capabilities of our framework, we train three different networks with various attacks: 1) Gaussian-Trained-Network (GT-Net) is trained under Gaussian noise attack ($\sigma$=3), 2) JPEG-Trained-Network (JT-Net) is trained under JPEG attack (quality=70), 3) Multi-Attack-Trained-Network (MT-Net) is trained in the presence of multiple attacks with equal probabilities, including salt \& pepper (4\%), Gaussian noise ($\sigma$=3), JPEG (quality=70), and mean smoothing filter (3$\times$3). 

Implementation details and the network configuration are discussed in \ref{sec:netconfig}. Training strategies and working with the trained networks are discussed in \ref{sec:expsetup}. We analyze the imperceptibility and robustness of the trained networks under several attacks in \ref{sec:expquantity} and \ref{sec:expquality}, then compare the results to the state-of-the-art watermarking algorithms in \ref{sec:compare}.  Finally, some watermarking patterns of the trained networks and their technical characteristics are discussed in \ref{sec:expanalysis}. The source codes of the framework will be uploaded on Github shortly. 

\subsection{Network Configurations}
\label{sec:netconfig}
In our experiments, we set the block size to $8\times 8$ and use training image patches of size $32\times 32$. So we have 16 blocks per input image which is used for embedding a $4\times 4$ watermark pattern. We use the DCT transform domain in our experiments, however any other linear transform may be applied. The embedding network (as shown in Fig. \ref{fig:embedding}) consists of two interface-transform-layers implemented by $1\times 1$ convolutional masks to perform the change-of-basis DCT operation and inverse DCT transform. The embedding network between the transform layers is composed of one $1\times 1$ convolution and four $2\times 2$ circular convolutional layers with Exponential Linear Unit (ELU) activation \cite{ELU}. All the layers of embedding network contain 64 convolution filters. The basic structure of the extraction network is very similar to the embedding module.  As demonstrated in Fig. \ref{fig:embedding}, the first layer is a transform layer composed of 64 DCT filter masks ($1\times 1$ convolutions). Then we have another $1\times 1$ convolution layer and three $2\times 2$ convolution layers with 64 filters per layer and ELU activation functions. A final $1\times 1$ convolutional layer consisting of one neuron with sigmoid activation generates the watermark probability map. A threshold is applied to this output to produce the final watermark. Strides of all filters in embedding and extraction modules are set to one. Width and height of images throughout the network are constant, as a result of using circular convolution.

\begin{table}[]
  \centering
  \caption{\small The network and training parameters}  
  
  \label{tab:parameters}
  \begin{adjustbox}{width=0.8\columnwidth}
  
  \begin{small}
	  \begin{tabular} {ccc}
	  \hline 
	    {\bf }          Parameter & Value & description  \\ 	
	  \hline
      (M,N)  &(8,8)  & Size of network blocks \\			
      (W,H)  &(32,32) & Size of training image patches  \\			
      (w,h)   &(4,4)    & Size of watermark \\			
      Iteration No.  &1000000	  & Training iteration number \\			
      LR & $10^{-4}$	& Learning rate \\ 			
      mo   &0.98 & Momentum of optimization \\			
      $\gamma$  & 0.75 for GT-NET and JT-NET,  & Relative loss function weights  \\	
      & 0.5 for MT-NET & \\			
	  \hline
	  \end{tabular}
  \end{small}  
  \end{adjustbox}
  \label{tab:parameters}
\end{table}

\subsection{Experimental Setup}
\label{sec:expsetup}
For training process, CIFAR10 \cite{CIFAR} and Pascal VOC2012 \cite{Pascal} datasets are combined to shape our training set of cover images. CIFAR10 consists of 60000 tiny RGB images ($32\times 32$), which is divided to training/test sets of size 50000/10000 images.
%CIFAR10 consists of 60000 tiny RGB images ($32\times 32$), from which we convert 50000 images of its training set 
We combine the 50000 CIFAR10 training set images with Pascal dataset and convert them to grayscale to be used as cover images for training. Since the Pascal dataset contains large, high-resolution images, we extract $32\times 32$ patches from random positions of dataset images. 
We want our dataset to contain a range of smooth to high-frequency patterns for better training.
Hence, among the set of generated patches we select a subset which equally contains all intensity variances. 
% uniform distribution of intensity variances. 
The final training set is a combination of two datasets containing around 334K grayscale image patches. In the training phase, we assign random watermarks to image patches in every iteration. In this way, we avoid biasing the network with a specific watermark pattern. We utilize the stochastic gradient descent algorithm for optimization and training. Some training configuration and parameters are shown in Table \ref{tab:parameters}. The training time for single attack networks is about 9 hours with 1000,000 iterations. For multi-attack training, the number of iterations is doubled and consequently, its training is twice slower than the training of the single attack network. 

\begin{table*}[!tb]
%\begin{table*}[!t]
\centering
\caption{\small Robustness and imperceptibility results of our networks.}
\label{lbl-table2}
\begin{adjustbox}{width=\textwidth}
\begin{tabular}{cc|c|c|ccc|ccc|ccc|ccc|ccc}
\hline
&&\multicolumn{2}{c|}{\textbf{Imperceptibility}}&\multicolumn{15}{c}{\textbf{Robustness (\% BER)}}\\
\hline
&\multirow{2}{*}{$\alpha$\hspace{2pt}} & \multirow{2}{*}{\textbf{\begin{tabular}[c]{@{}c@{}}PSNR\\ (dB)\end{tabular}}} & \multirow{2}{*}{\textbf{SSIM}} & \multicolumn{3}{c|}{\textbf{\begin{tabular}[c]{@{}c@{}}Gaussian Noise \\ ($\sigma$)\end{tabular}}} & \multicolumn{3}{c|}{\textbf{\begin{tabular}[c]{@{}c@{}}Salt \& \\ Pepper (\%)\end{tabular}}} & \multicolumn{3}{c|}{\textbf{\begin{tabular}[c]{@{}c@{}}Cropping\\ (\%)\end{tabular}}} & \multicolumn{3}{c|}{\textbf{\begin{tabular}[c]{@{}c@{}}Grid Crop\\ (\%)\end{tabular}}} & \multicolumn{3}{c}{\textbf{\begin{tabular}[c]{@{}c@{}}patterned-pixel- \\elimination (lines)\end{tabular}}} \\ \cline{5-19} 
\multicolumn{2}{c|}{} &  &  & \textbf{5} & \textbf{15} & \textbf{25} & \textbf{2} & \textbf{6} & \textbf{10} & \textbf{10} & \textbf{20} & \textbf{30} & \textbf{20} & \textbf{30} & \textbf{40} & \textbf{3} & \textbf{6} & \textbf{9} \\ \hline
\multirow{3}{*}{\textbf{MT-Net}} & \textbf{1.0} & 35.93 & 0.966 & 0.0 & 2.6 & 12.7 & 0.0 & 0.1 & 0.9 & 6.0 & 11.3 & 17.1 & 3.3 & 6.9 & 11.4 & 1.4 & 2.6 & 5.0 \\
 & \textbf{0.8} & 37.84 & 0.978 & 0.1 & 6.4 & 18.2 & 0.1 & 0.6 & 2.7 & 6.0 & 11.5 & 17.1 & 4.4 & 8.5 & 13.4 & 1.9 & 3.4 & 6.5 \\
 & \textbf{0.6} & 40.24 & 0.987 & 2.4 & 14.5 & 25.6 & 2.9 & 4.5 & 9.1 & 7.7 & 13.1 & 18.8 & 8.3 & 13.0 & 18.0 & 4.0 & 6.3 & 10.3 \\ \hline
\multirow{3}{*}{\textbf{JT-Net}} & \textbf{1.0} & 39.77 & 0.985 & 0.6 & 17.0 & 27.7 & 14.9 & 30.9 & 36.3 & 5.8 & 11.3 & 16.8 & 29.3 & 35.7 & 39.4 & 3.3 & 5.4 & 9.9 \\
 & \textbf{0.8} & 41.53 & 0.990 & 4.3 & 22.6 & 31.7 & 20.5 & 34.5 & 38.9 & 7.2 & 12.6 & 18.3 & 32.8 & 38.2 & N & 5.0 & 7.6 & 12.5 \\
 & \textbf{0.6} & 43.66 & 0.994 & 13.4 & 29.2 & 36.1 & 27.6 & 38.3 & N & 13.3 & 18.1 & 22.9 & 37.0 & N & N & 11.5 & 14.5 & 19.2 \\ \hline
\multirow{3}{*}{\textbf{GT-Net}} & \textbf{1.0} & 44.14 & 0.992 & 0.5 & 17.8 & 28.9 & 14.6 & 31.6 & 37.1 & 5.6 & 10.9 & 16.5 & 3.4 & 6.8 & 11.2 & 0.6 & 1.5 & 3.2 \\
 & \textbf{0.8} & 45.73 & 0.994 & 2.4 & 23.0 & 32.6 & 19.1 & 34.8 & 39.5 & 5.7 & 11.1 & 16.6 & 3.7 & 7.2 & 11.6 & 0.9 & 2.1 & 4.2 \\
 & 0.6 & 47.52 & 0.996 & 8.0 & 28.8 & 36.6 & 25.0 & 38.4 & N & 7.5 & 12.8 & 18.3 & 6.1 & 9.7 & 14.0 & 3.0 & 4.5 & 7.2 \\ \hline
\end{tabular}
\end{adjustbox}
\label{tab:long1}
%\end{table*}
\vspace{5pt}
%\begin{table*}[]
\centering
\caption{\small Robustness and imperceptibility results of our networks. (Continued)}
\label{lbl-table3}
\begin{adjustbox}{width=\textwidth}
\begin{tabular}{cc|c|c|ccc|ccc|ccc|ccc|ccc}
\hline
&&\multicolumn{2}{c|}{\textbf{Imperceptibility}}&\multicolumn{15}{c}{\textbf{Robustness (\% BER)}}\\
\hline
&\multirow{2}{*}{$\alpha$\hspace{2pt}} & \multirow{2}{*}{\textbf{\begin{tabular}[|c]{@{}c@{}}PSNR\\ (dB)\end{tabular}}} & \multirow{2}{*}{\textbf{SSIM}} & \multicolumn{3}{c|}{\textbf{JPEG (quality)}} & \multicolumn{3}{c|}{\textbf{\begin{tabular}[c]{@{}c@{}}Gaussian Blur\\ (radius)\end{tabular}}} & \multicolumn{3}{c|}{\textbf{\begin{tabular}[c]{@{}c@{}}Sharpening \\  (radius)\end{tabular}}} & \multicolumn{3}{c|}{\textbf{\begin{tabular}[c]{@{}c@{}}Median\\ (window size)\end{tabular}}} & \multicolumn{3}{c}{\textbf{\begin{tabular}[c]{@{}c@{}}Resizing \\ (scale)\end{tabular}}} \\ \cline{5-19} 
\multicolumn{2}{c|}{} &  &  & \textbf{90} & \textbf{70} & \textbf{50} & \textbf{1} & \textbf{1.6} & \textbf{2} & \textbf{1} & \textbf{5} & \textbf{10} & \textbf{3} & \textbf{5} & \textbf{7} & \textbf{0.5} & \textbf{0.75} & \textbf{1.5} \\ \hline
\multirow{3}{*}{\textbf{MT-Net}} & \textbf{1.0} & 35.93 & 0.966 & 0.0 & 0.0 & 1.2 & 0.1 & 29.0 & N & 0.0 & 0.1 & 0.2 & 2.9 & N & N & 5.9 & 0.0 & 0.0 \\
 & \textbf{0.8} & 37.84 & 0.978 & 0.0 & 0.4 & 3.5 & 1.6 & 34.3 & N & 0.1 & 0.3 & 0.6 & 6.1 & N & N & 12.3 & 0.3 & 0.1 \\
 & \textbf{0.6} & 40.24 & 0.987 & 1.6 & 4.2 & 11.8 & 8.6 & 39.0 & N & 0.9 & 2.4 & 3.2 & 13.4 & N & N & 21.2 & 3.8 & 2.6 \\ \hline
\multirow{3}{*}{\textbf{JT-Net}} & \textbf{1.0} & 39.77 & 0.985 & 0.1 & 0.2 & 1.3 & 24.7 & N & N & 0.8 & 1.6 & 2.1 & 28.9 & N & N & 28.7 & 8.8 & 6.3 \\
 & \textbf{0.8} & 41.53 & 0.990 & 1.3 & 2.3 & 6.7 & 28.7 & N & N & 1.4 & 3.4 & 4.4 & 31.7 & N & N & 32.0 & 14.5 & 12.2 \\
 & \textbf{0.6} & 43.66 & 0.994 & 7.9 & 10.7 & 17.3 & 32.9 & N & N & 5.6 & 9.1 & 10.4 & 34.3 & N & N & 35.0 & 21.0 & 19.2 \\ \hline
\multirow{3}{*}{\textbf{GT-Net}} & \textbf{1.0} & 44.14 & 0.992 & 9.6 & N & N & N & N & N & 0.2 & 0.9 & 1.4 & N & N & N & N & 37.9 & 1.6 \\
 & \textbf{0.8} & 45.73 & 0.994 & 16.0 & N & N & N & N & N & 0.6 & 1.6 & 2.3 & N & N & N & N & 39.3 & 3.2 \\
 & 0.6 & 47.52 & 0.996 & 24.2 & N & N & N & N & N & 2.5 & 3.8 & 4.7 & N & N & N & N & N & 7.2 \\ \hline
\end{tabular}
\end{adjustbox}
\label{tab:long2}
\end{table*}
%#################################
For evaluating the trained networks, we embed 1024 bit ($32 \times 32$) watermarks in $512 \times 512$ gray-scale images. The watermark is embedded with four times redundancy, as the cover image contains $64 \times 64$ image blocks. For this purpose, a $64\times 64$ bit redundant plane is formed so that each bit of watermark is repeated four times in a regular pattern in this plane. We refer to it as watermark plane. Then the cover image is partitioned into $32\times 32$ sub-images, and the watermark plane is partitioned into $4\times 4$ sub-watermarks for feeding to the embedding network. The embedding network produces watermarked sub-images which are tiled with each other to form the watermarked image. The watermarked image is then passed through several attacks to simulate the real world situation. For the extraction phase, we follow the same protocol. First, the attacked watermarked image is partitioned into $32\times 32$ sub-images. These sub-images are fed into the extraction network, which extracts $4\times 4$ patches of the watermark plane. Then these patches are tiled to form a $64\times 64$ redundant watermark plane. Finally, a voting procedure is applied on the corresponding bits to produce the 1024-bit watermark data.

\subsection{Quantitative Results}
\label{sec:expquantity}
To analyze the trained networks (GT-Net, JT-Net, and MT-Net), we test them on all 49 images in the Granada dataset \cite{dataset}. The imperceptibility of each watermarked image is presented in terms of PSNR and SSIM. Tables \ref{tab:long1} and \ref{tab:long2} demonstrate the PSNR and SSIM of the watermarked images produced by all the trained networks with three different strength factors ($\alpha$). PSNR and SSIM of a single image are calculated by averaging 20 watermarked images with different random watermarks. We follow the same process for all the 49 images of Granada dataset and present their average values as the network performance. To demonstrate the robustness of the proposed networks, BER of extracted watermarks under several attacks are calculated for three different strength factors. Similar to PSNR and SSIM, all the numeric results of both tables represent the average result over all images in the test dataset with 20 random watermarks per image.

Robustness of the networks is tested for three levels of each attack. The symbol N in the tables stands for Non-robust and is used when the BER value is around $50\%$. Gaussian noise is applied with three different standard deviations. The parameter in salt and pepper and cropping attacks is the percentage of changed pixels. To show the strength of the proposed networks in diffusion watermarking and data sharing, a new attack is introduced called Grid cropping, in which random $8 \times 8$ blocks throughout the image are suppressed to zero. For the Gaussian blur and sharpening (/unmask) attacks, the attack parameter represents the filter's radius. The Median filter parameter demonstrates the filter mask size. In resizing attack, the image is resized by the shown scale and resized back to original image based on bilinear interpolation. Similar to cropping and salt and pepper, we conduct the grid cropping attack in three different levels, representing the percentage of suppressed blocks. The BER results in \ref{tab:long1} declare that even if a meaningful number of image blocks are cropped, we can still extract the majority of the watermark data. The next rarely used attack is the patterned-pixel-elimination attack, in which a text is written on the watermarked image. The attack parameter represents the number of text lines in ``Natural script'' handwriting with the font size of 40. Similar to Grid cropping, this attack highlights the network's capability in diffusing the watermark data throughout the cover image. Tables \ref{tab:long1} and \ref{tab:long2} demonstrate that the MT-NET exhibits overall better robustness compared to the other two networks, at the expense of lower PSNR and SSIM as expected. This is similar in spirit to a general belief in Multi-Task Learning (MTL) \cite{p49}, which states training one neural network for multiple similar tasks leads to overall better performance compared to training separate networks for every single task. As shown in Tables \ref{tab:long1} and \ref{tab:long2}, regardless of some exceptions, MT-NET demonstrates better extraction rate (lower BER) than the other two networks. This is even true for the attacks that JT-Net and GT-Net are trained for (JPEG attack and Gaussian noise attack).
\begin{figure}[!t]
  \centering
	  \begin{minipage}{1.9\textwidth}	  
			% ==========================================
		  \begin{minipage}{1.0\columnwidth}
		  	\begin{minipage}{0.05\columnwidth}
		  	  \centering
		  		 \textcolor{white}{ .}
		  	\end{minipage}
		  	\begin{minipage}{0.225\columnwidth}
		  	  \centering
		  		{\scriptsize PSNR}
		  	\end{minipage}
		  	\begin{minipage}{0.225\columnwidth}
		  	  \centering
		  		{\scriptsize SSIM}
		  	\end{minipage}
		  \end{minipage}
		  \begin{minipage}{1.0\columnwidth}
		  	\begin{minipage}{0.04\columnwidth}
		  	  \centering
		  		\begin{scriptsize}
		  			GT-NET
		  		\end{scriptsize}
		  	\end{minipage}
		  	\begin{minipage}{0.225\columnwidth}
		  	  \includegraphics[width=0.95\columnwidth]{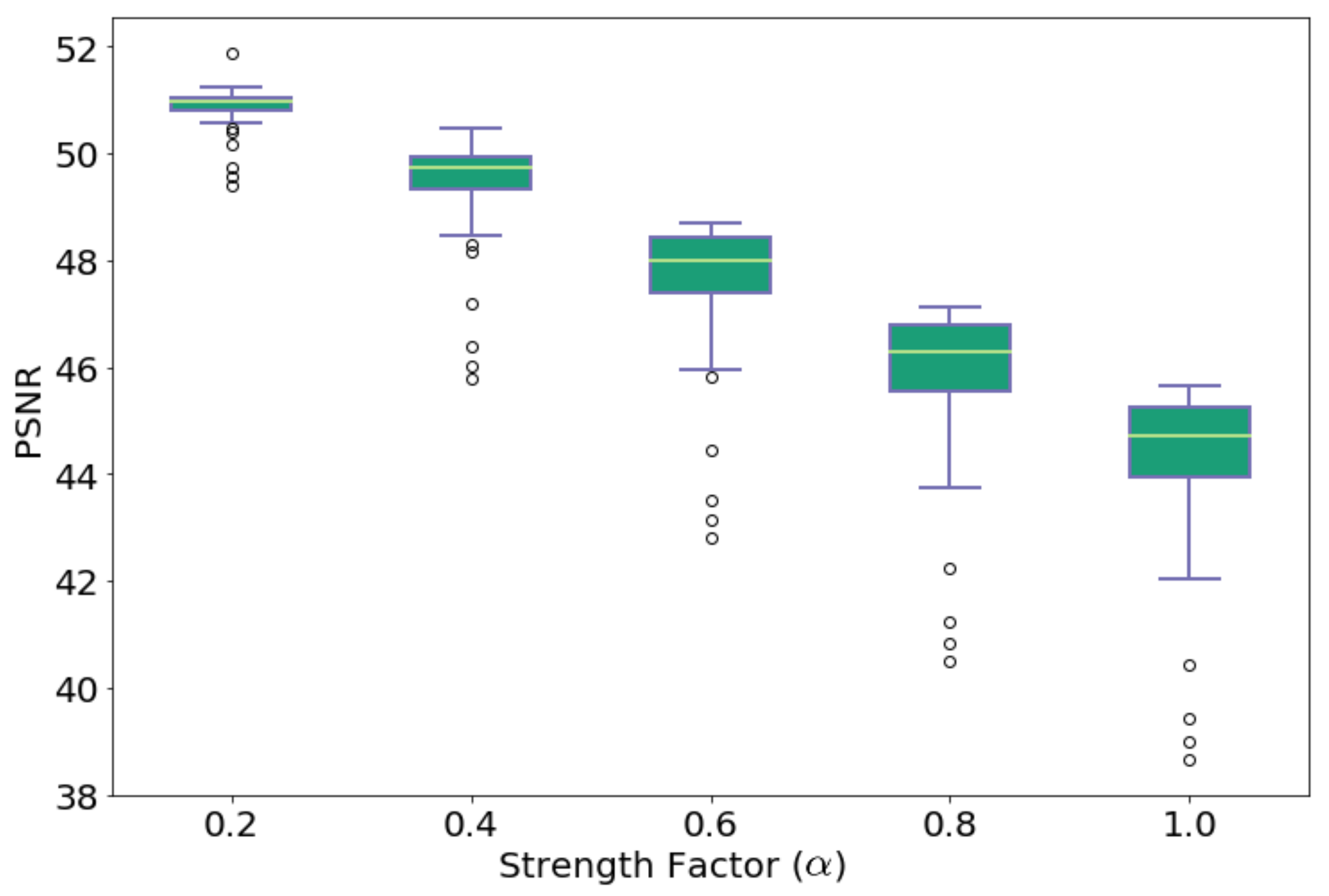}
		  	\end{minipage}
		  	\begin{minipage}{0.225\columnwidth}
		  	  \includegraphics[width=1.0\columnwidth]{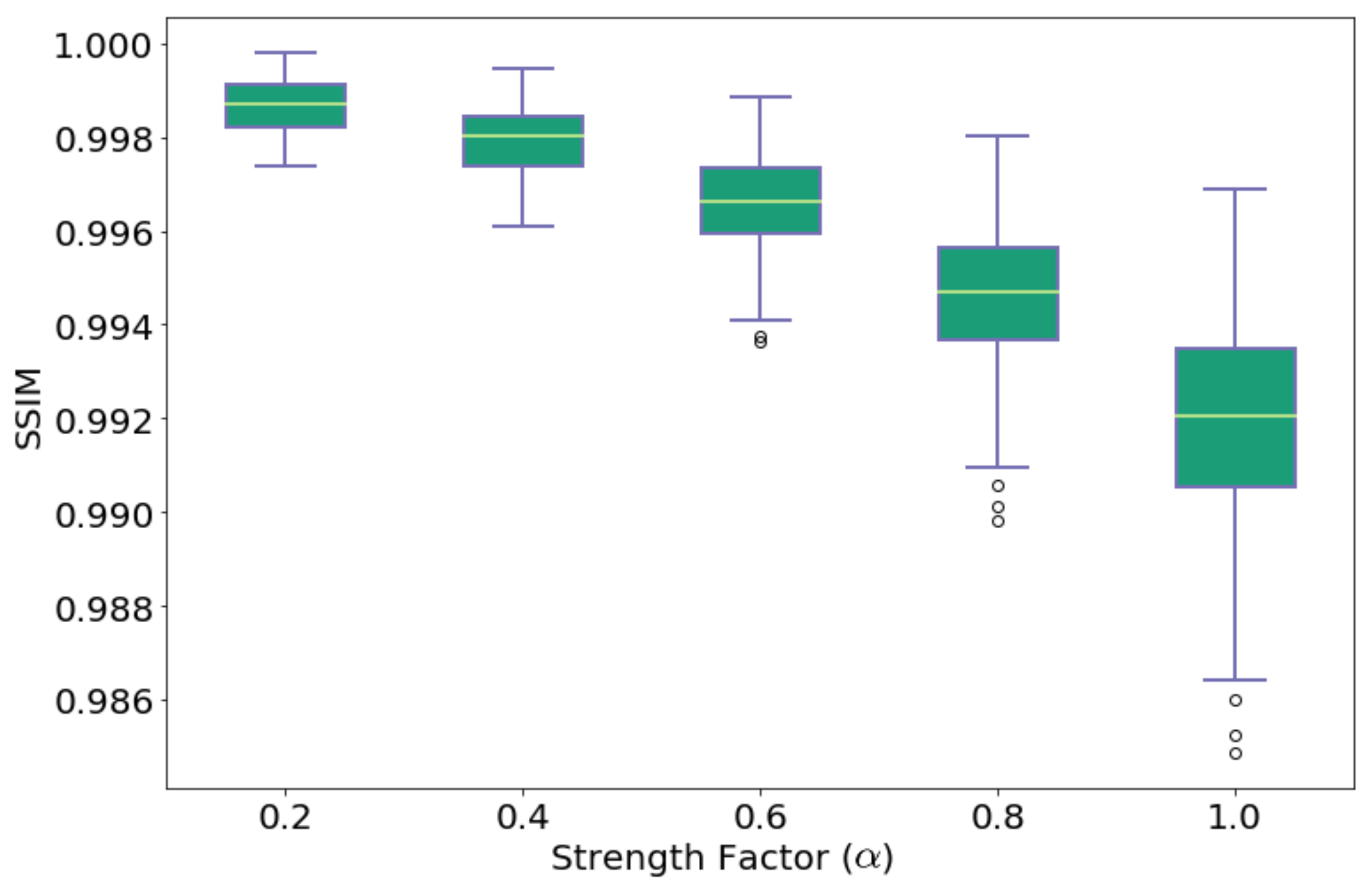}
		  	\end{minipage}
		  \end{minipage}
		  \begin{minipage}{1.0\columnwidth}
		  	\begin{minipage}{0.04\columnwidth}
		  	  \centering
		  		\begin{scriptsize}
		  			JT-NET
	  			\end{scriptsize}
		  	\end{minipage}
		  	\begin{minipage}{0.225\columnwidth}
		  		\includegraphics[width=0.95\columnwidth]{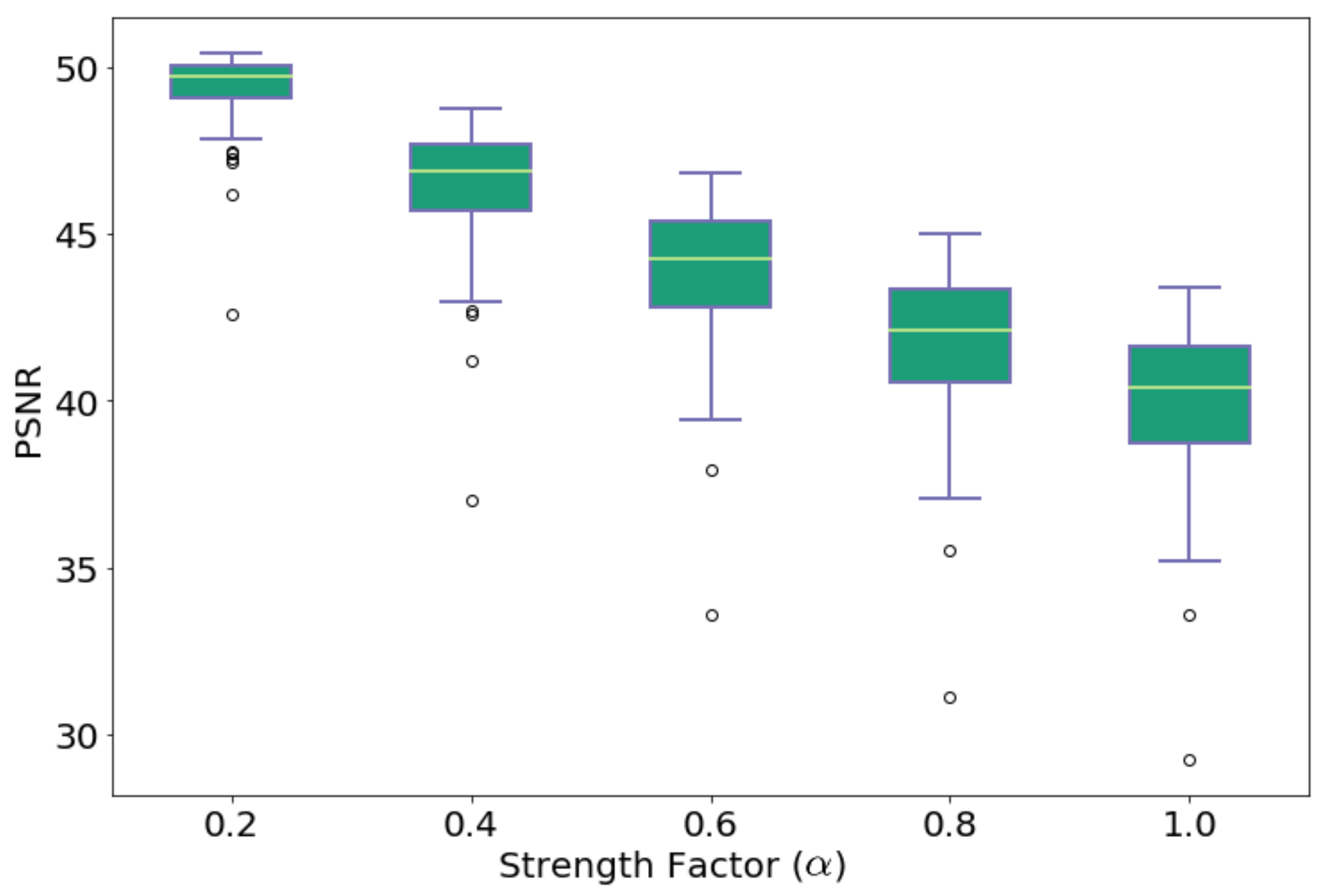} 
		  	\end{minipage}
		  	\begin{minipage}{0.225\columnwidth}
			  	\includegraphics[width=1.0\columnwidth]{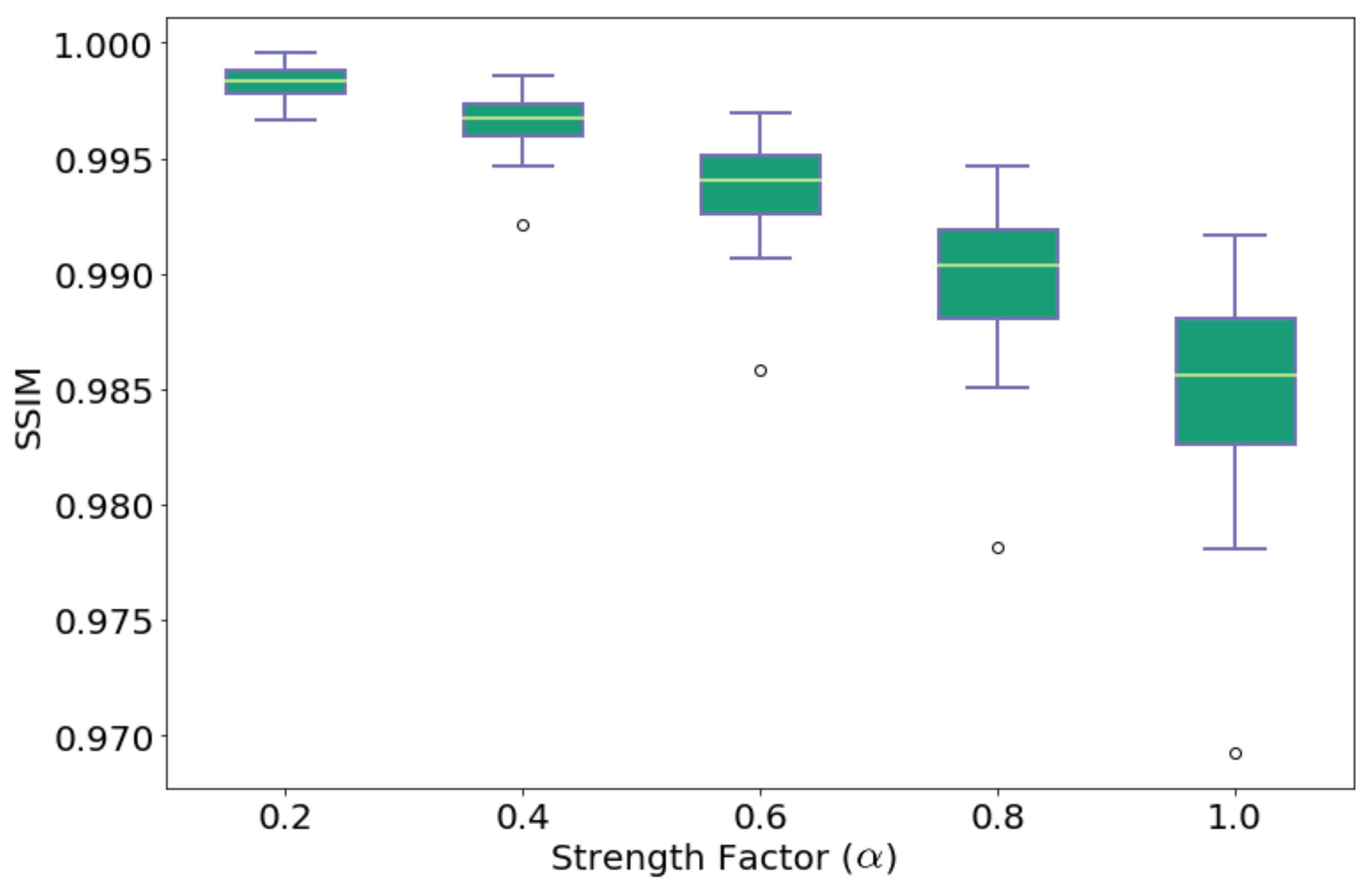} 
		  	\end{minipage}
		  \end{minipage}
		  \begin{minipage}{1.0\columnwidth}
		  	\begin{minipage}{0.04\columnwidth}
		  	  \centering
		  	  \begin{scriptsize}
		  	  	MN-NET
		  	  \end{scriptsize}
		  	\end{minipage}
		  	\begin{minipage}{0.225\columnwidth}
			  	\includegraphics[width=0.95\columnwidth]{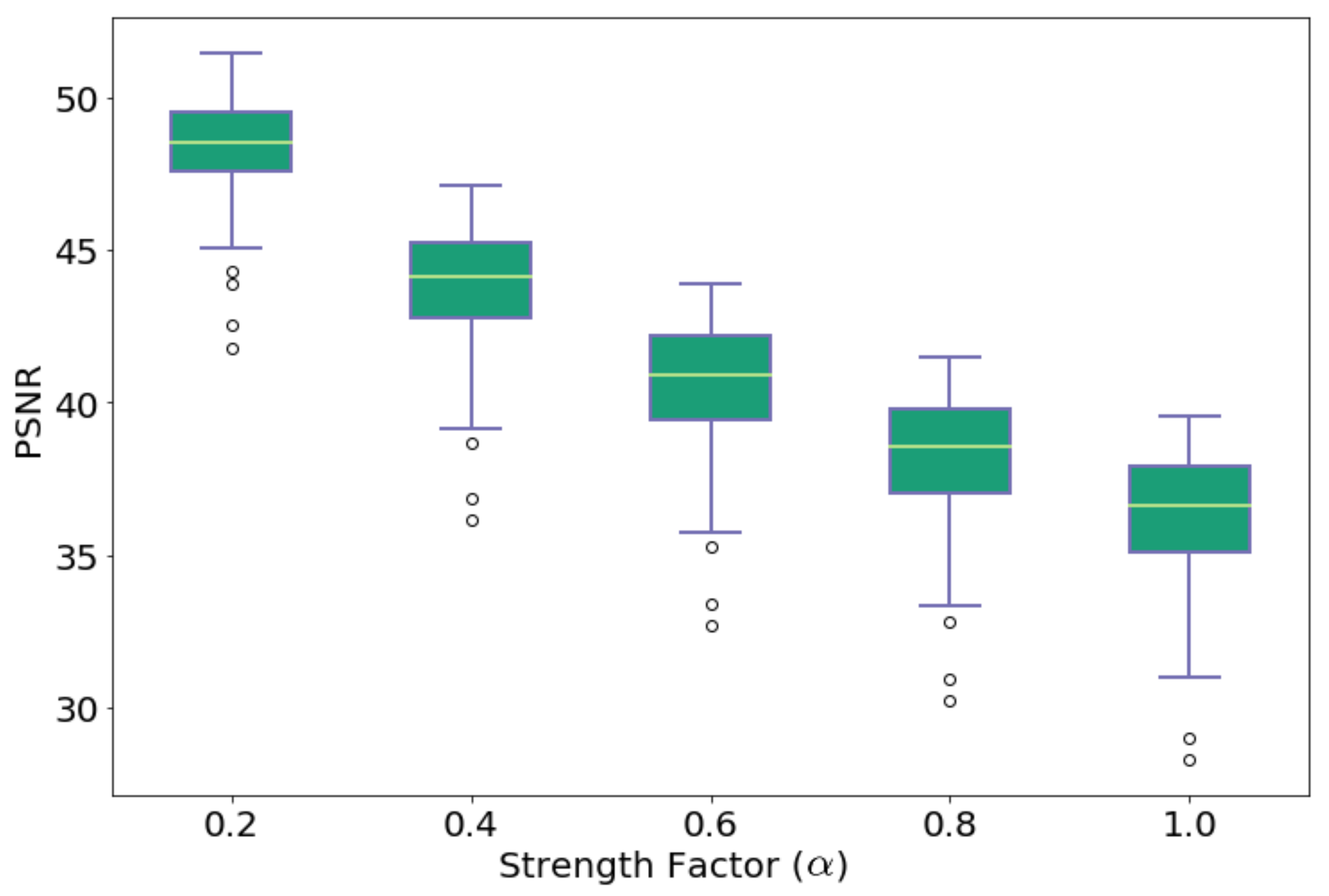} 
		  	\end{minipage}
		  	\begin{minipage}{0.225\columnwidth}
			  	\includegraphics[width=1.0\columnwidth]{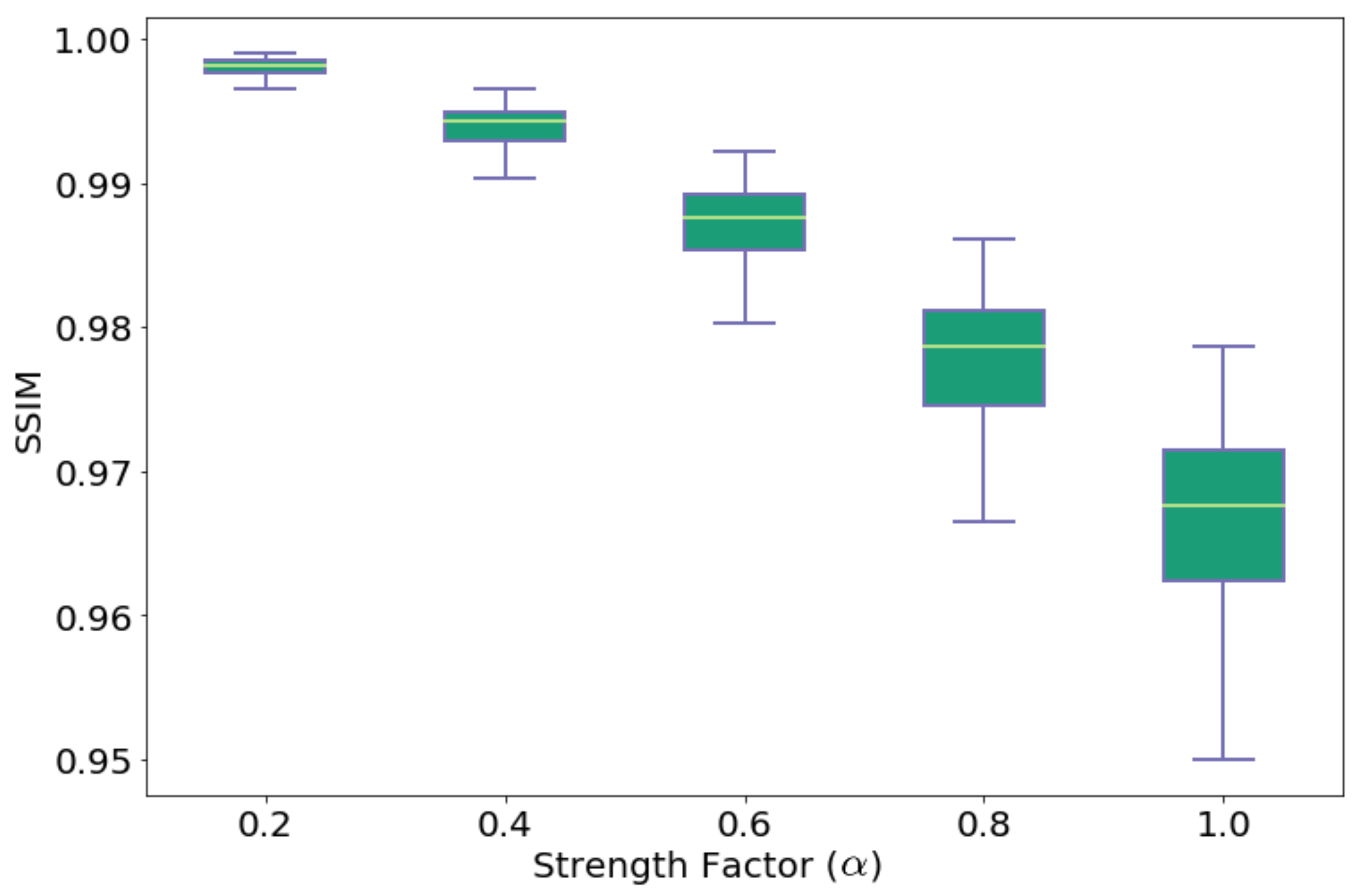} 
		  	\end{minipage}
		  \end{minipage}
		  % ==========================================
	  \end{minipage}
	
 \caption
    {\small 
box-plot representation of PSNR and SSIM for the trained networks. Each figure illustrates the variations of PSNR/SSIM versus $\alpha$. 		
    }
  \label{fig:boxplot}
\end{figure}
Another important characteristic of the system is that the robustness of the watermarking networks can be controlled by two means. In the training mode, when the network confronts more powerful attacks, the embedding module embeds stronger watermark symbols in the image. This strong embedding leads to more robustness at the expense of lower PSNR. In spite of that, in a trained network we can still control this trade-off by tuning the watermarking strength factor. As shown in Tables \ref{tab:long1} and \ref{tab:long2}, for a given attack and a fixed network, increasing the strength factor ($\alpha$) results in lower BER values. 

For further details about the network behavior, box-plots of Fig. \ref{fig:boxplot} display the range and variations of PSNR and SSIM for all the watermarked images. Tables \ref{tab:long1} and \ref{tab:long2} only display the average outcomes of the proposed algorithm on the Granada dataset. In Fig. \ref{fig:boxplot}, each box corresponds to a strength factor and shows the distribution of SSIM and PSNR for all of the test images. Each row in Fig. \ref{fig:boxplot} is for one of the proposed networks. Increasing the strength factor lowers PSNR, and SSIM values.

\subsection{Qualitative Results}
\label{sec:expquality}
\begin{figure*}[]
\centering
\vspace{-75pt}
\begin{subfigure}{.32\textwidth}
  \centering
  \includegraphics[width=.99\linewidth]{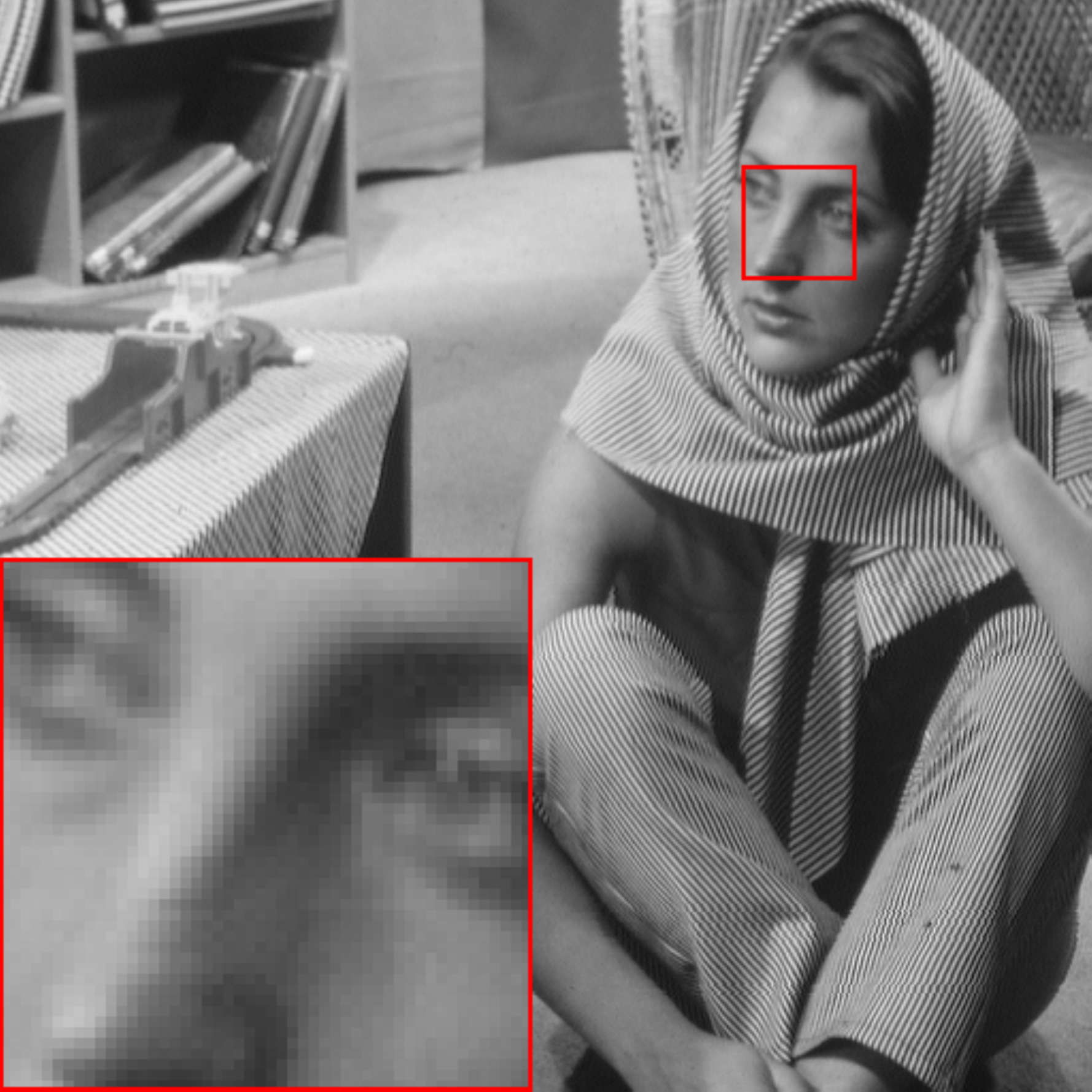}
  \caption{Cover Image}
  \label{fig:sfig_cover}
\end{subfigure}%
% Second Row
\vspace{6pt}
\begin{subfigure}{1.0\textwidth}
\centering
	\begin{subfigure}{.32\textwidth}
  		\centering
 	 	\includegraphics[width=.99\linewidth]{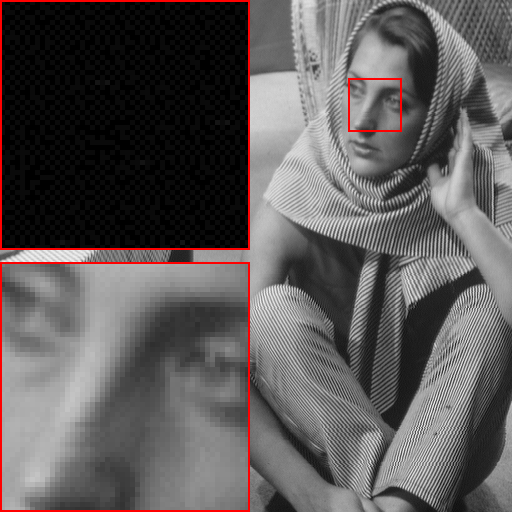}
  		\caption{GT-Net, $\alpha$=0.2}
  		\label{fig:sfig_gaussian02}
	\end{subfigure}
	\begin{subfigure}{.32\textwidth}
  		\centering
 	 	\includegraphics[width=.99\linewidth]{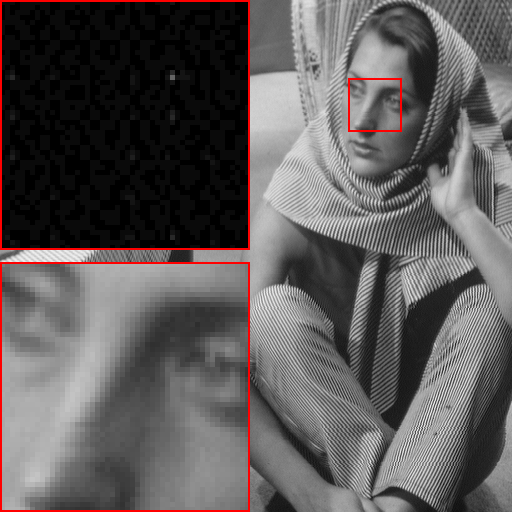}
  		\caption{JT-Net, $\alpha$=0.2}
  		\label{fig:sfig_jpeg02}
	\end{subfigure}
	\begin{subfigure}{.32\textwidth}
  		\centering
 	 	\includegraphics[width=.99\linewidth]{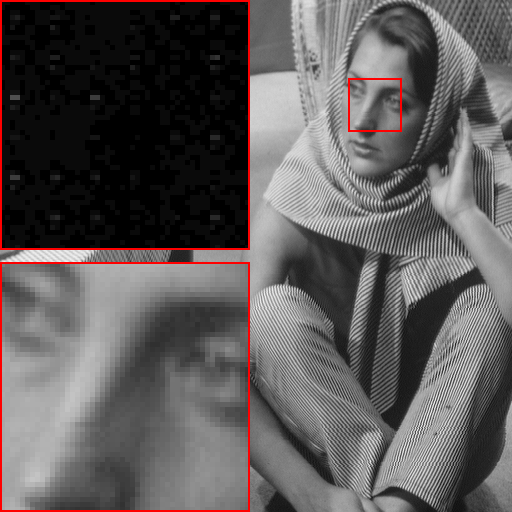}
  		\caption{MT-Net, $\alpha$=0.2}
  		\label{fig:sfig_multi02}
	\end{subfigure}
\end{subfigure}

% Third Row
\vspace{6pt}
\begin{subfigure}{1.0\textwidth}
\centering
	\begin{subfigure}{.32\textwidth}
  		\centering
 	 	\includegraphics[width=.99\linewidth]{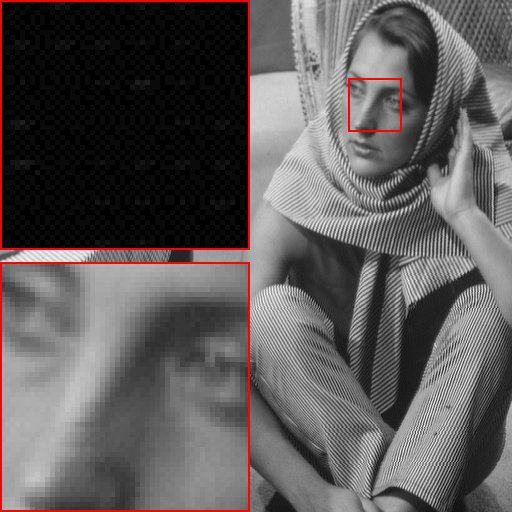}
  		\caption{GT-Net, $\alpha$=0.4}
  		\label{fig:sfig_gaussian04}
	\end{subfigure}
	\begin{subfigure}{.32\textwidth}
  		\centering
 	 	\includegraphics[width=.99\linewidth]{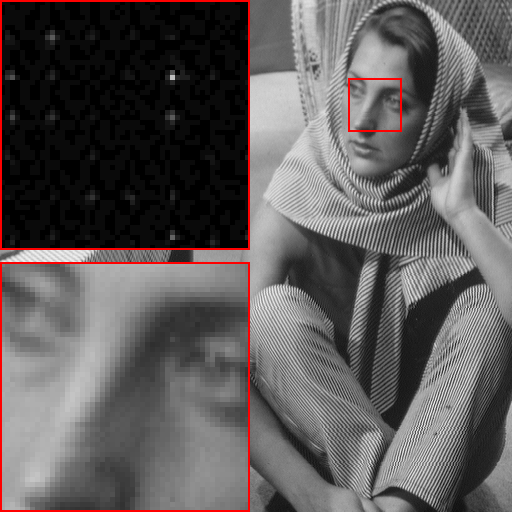}
  		\caption{JT-Net, $\alpha$=0.4}
  		\label{fig:sfig_jpeg04}
	\end{subfigure}
	\begin{subfigure}{.32\textwidth}
  		\centering
 	 	\includegraphics[width=.99\linewidth]{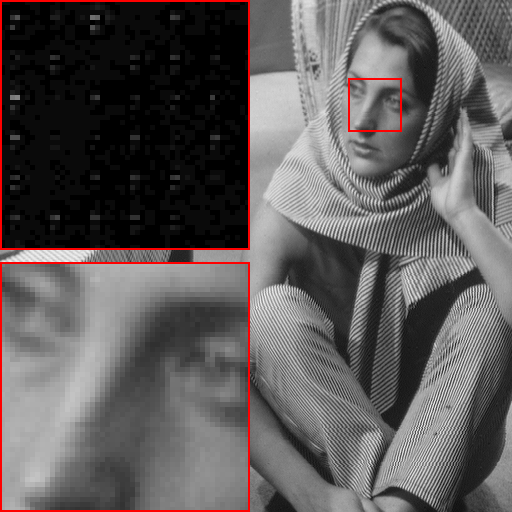}
  		\caption{MT-Net, $\alpha$=0.4}
  		\label{fig:sfig_multi04}
	\end{subfigure}
\end{subfigure}

% Forth Row
\vspace{6pt}
\begin{subfigure}{1.0\textwidth}
\centering
	\begin{subfigure}{.32\textwidth}
  		\centering
 	 	\includegraphics[width=.99\linewidth]{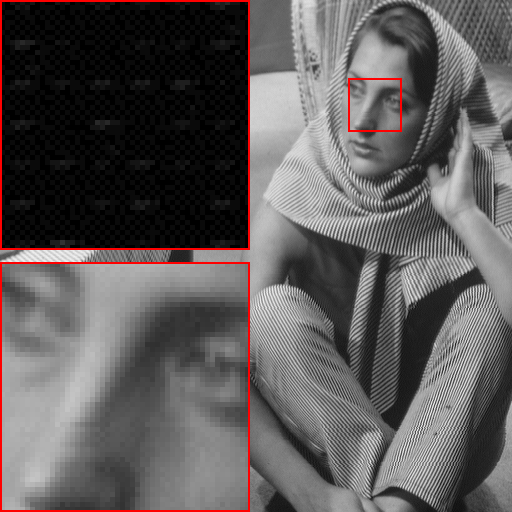}
  		\caption{GT-Net, $\alpha$=0.6}
  		\label{fig:sfig_gaussian04}
	\end{subfigure}
	\begin{subfigure}{.32\textwidth}
  		\centering
 	 	\includegraphics[width=.99\linewidth]{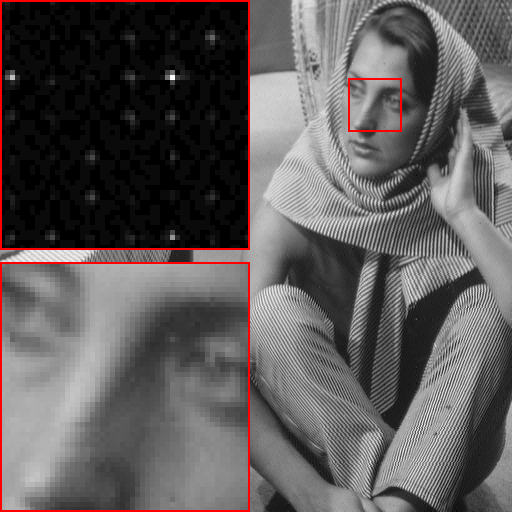}
  		\caption{JT-Net, $\alpha$=0.6}
  		\label{fig:sfig_jpeg04}
	\end{subfigure}
	\begin{subfigure}{.32\textwidth}
  		\centering
 	 	\includegraphics[width=.99\linewidth]{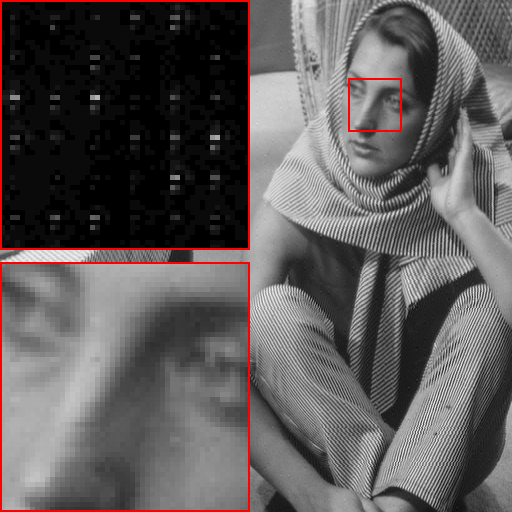}
  		\caption{MT-Net, $\alpha$=0.6}
  		\label{fig:sfig_multi04}
	\end{subfigure}
\end{subfigure}
\vspace{-6pt}
\caption{\small The watermarked image produced by three networks for various strength factors. A square region of the watermarked image and its amplified absolute difference with original image are enlarged for visualization.}
\label{fig:watermarked}
\end{figure*}

For visualization purposes, the watermarked images produced by the three networks are illustrated in Fig. \ref{fig:watermarked}. In Fig. \ref{fig:watermarked} a random 128-bit watermark is embedded in Barbara image from the Granada dataset \cite{dataset} by using GT-Net, JT-Net, and MT-Net.  The absolute difference between the cover image and the watermarked image is illustrated to show the watermark pattern. For better visualization, this difference is multiplied by 10. Furthermore, a small area of the image is zoomed for better illustration.  We may notice in the difference images that amplitudes of the produced artifacts vary in different areas of the cover image. These variations in the difference matrices imply that the watermark symbols are adaptively embedded based on the local features of the image.

\begin{figure*}[]
\centering
% First Row
\begin{subfigure}{1.0\textwidth}
\centering
	\begin{subfigure}{.32\textwidth}
  		\centering
 	 	\includegraphics[width=.99\linewidth]{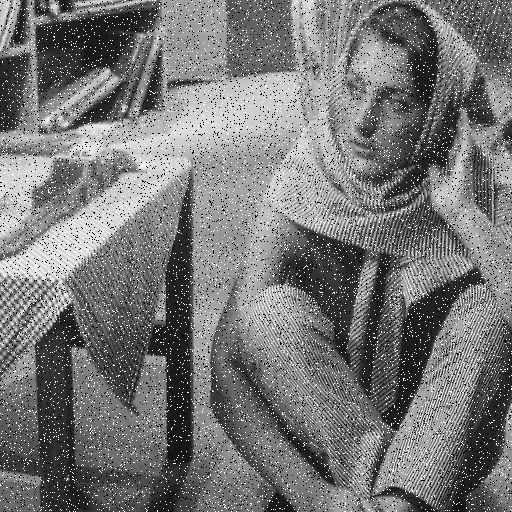}
  		\caption{Salt and pepper \\ (10\%) BER=8.8\%}
  		\label{fig:sfig_salt}
	\end{subfigure}
	\begin{subfigure}{.32\textwidth}
  		\centering
 	 	\includegraphics[width=.99\linewidth]{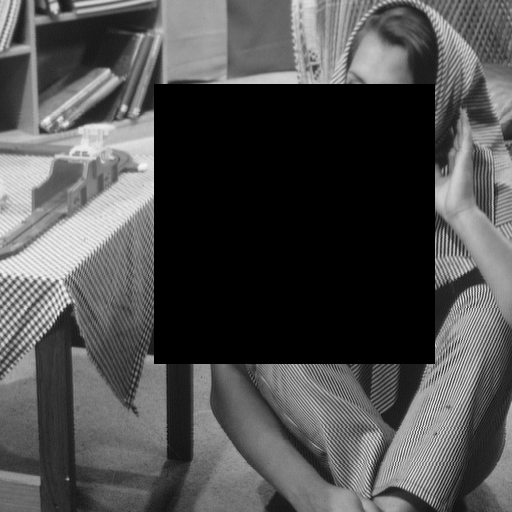}
  		\caption{Cropping (30\%) \\ BER=18.5\%}
  		\label{fig:sfig_crop}
	\end{subfigure}
	\begin{subfigure}{.32\textwidth}
  		\centering
 	 	\includegraphics[width=.99\linewidth]{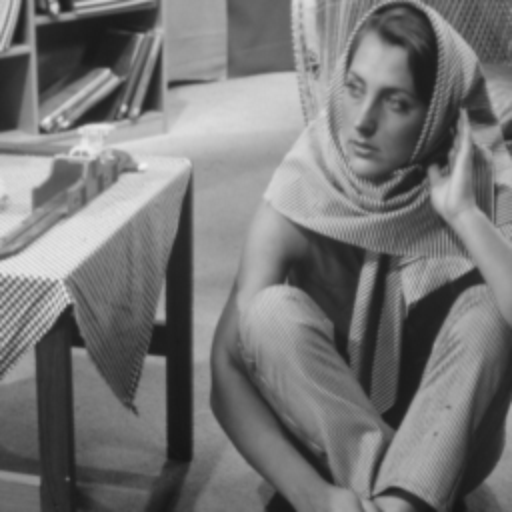}
  		\caption{Resizing (0.5) \\ BER=18.1\%}
  		\label{fig:sfig_resize}
	\end{subfigure}
\end{subfigure}

% Second Row
\vspace{10pt}
\begin{subfigure}{1.0\textwidth}
\centering
	\begin{subfigure}{.32\textwidth}
  		\centering
 	 	\includegraphics[width=.99\linewidth]{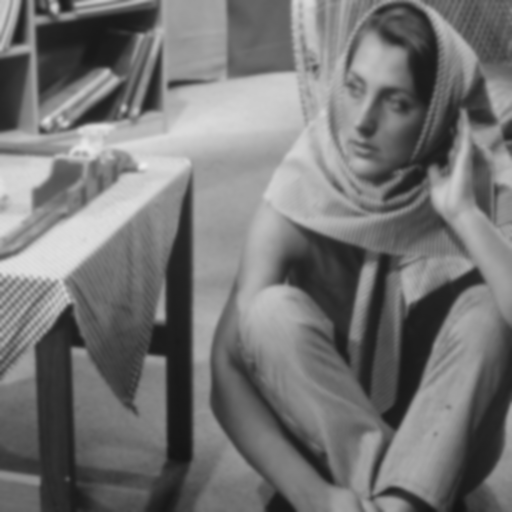}
  		\caption{Gaussian blur (Rad=1.6) \\ BER=36.3\%}
  		\label{fig:sfig_gaussianblur}
	\end{subfigure}
	\begin{subfigure}{.32\textwidth}
  		\centering
 	 	\includegraphics[width=.99\linewidth]{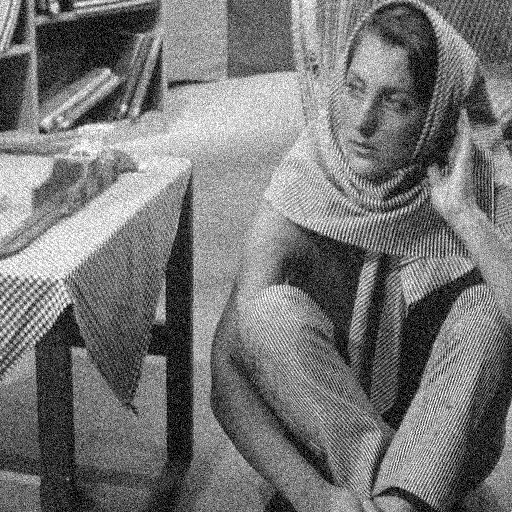}
  		\caption{Gaussian noise ($\sigma$=25) \\ BER=26.0\%}
  		\label{fig:sfig_gaussiannoise}
	\end{subfigure}
	\begin{subfigure}{.32\textwidth}
		\begin{center}
 	 	\includegraphics[width=.99\linewidth]{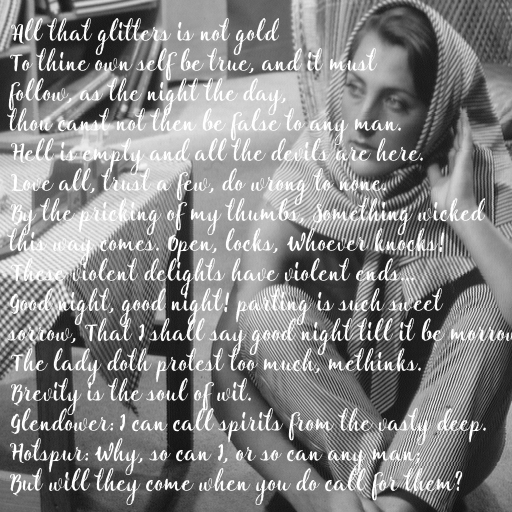}
  		\caption{\small Patterned Elimination \\ (lines=16) BER=18.0\%}
  		\label{fig:sfig_painting}
  		\end{center}
	\end{subfigure}
\end{subfigure}

% Third Row
\vspace{10pt}
\begin{subfigure}{1.0\textwidth}
\centering
	\begin{subfigure}{.32\textwidth}
  		\centering
 	 	\includegraphics[width=.99\linewidth]{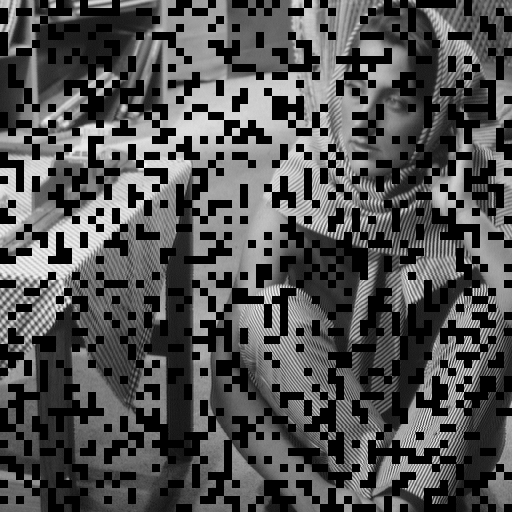}
  		\caption{Grid crop (30\%) \\ BER=13.4\%}
  		\label{fig:sfig_gridcrop}
	\end{subfigure}
	\begin{subfigure}{.32\textwidth}
  		\centering
 	 	\includegraphics[width=.99\linewidth]{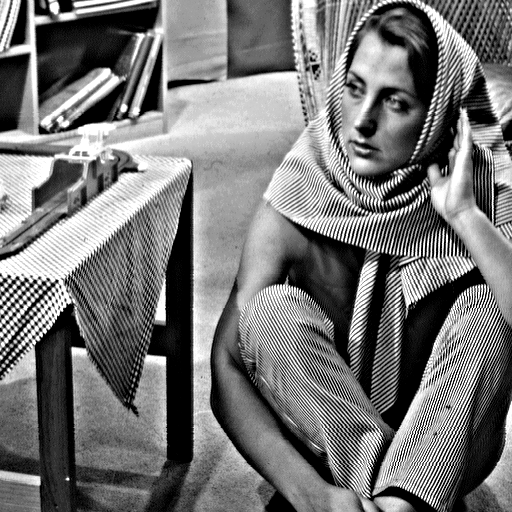}
  		\caption{Sharpenning (Rad=10) \\ BER=3.5\%}
  		\label{fig:sfig_sharpening}
	\end{subfigure}
	\begin{subfigure}{.32\textwidth}
  		\centering
 	 	\includegraphics[width=.99\linewidth]{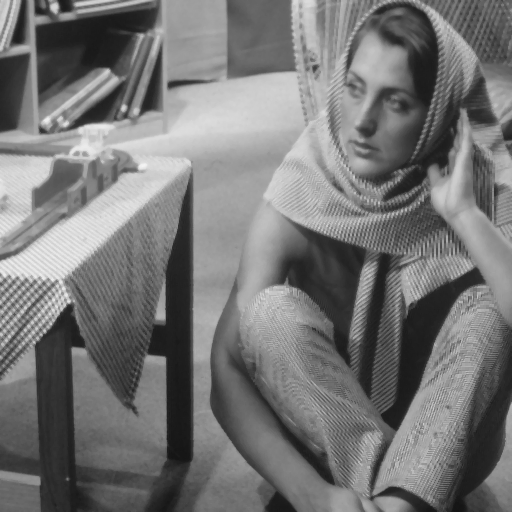}
  		\caption{\small Median window ($3\times 3$) \\ \small BER=10.3\%}
  		\label{fig:sfig_median}
	\end{subfigure}
\end{subfigure}

\caption{\small Visual effects of various attacks on Barbara. For each attack, the attack level/parameter and the resulting extraction BER are shown for MT-NET ($\alpha$=0.6)}
\label{fig:attacked}
\end{figure*}

Fig. \ref{fig:attacked} demonstrates some attacks on Barbara. For each attack, the attack level/parameter is displayed in parenthesis. The extraction BER for the attacked images are also reported using MT-NET with $\alpha=0.6$.

\subsection{Comparison With State-of-the-art}
\label{sec:compare}
In this section, we compare our network performance against \cite{Hidden} and \cite{p36}. 
The authors of HiDDeN framework \cite{Hidden} use COCO \cite{COCO} for their experiments. For a fair comparison, we match our testing conditions to HiDDeN, \ie, 30-bit random watermarks are embedded in $128\times 128$ color images. Hence, we redundantly embed in all channels of YUV space. In each channel, the parameter $\alpha$ is adjusted so that PSNR of both systems are similar.   Cropout and Dropout attacks are applied based on their definition. According to Table \ref{tab:Hidden}, although \paperName{} is not trained on cropout and dropout attacks, BER values under these two attacks are comparable to HiDDen.
\begin{table}[tb]
\renewcommand{\arraystretch}{1.5}
\centering
\caption{\small Comparison of MT-NET to network introduced in \cite{Hidden}}
\begin{adjustbox}{width=1.0\textwidth}
\begin{tabular}{llllll}
\hline
\multirow{2}{*}{\textbf{Method}} & \multicolumn{5}{c}{\textbf{Robustness (BER\%)}}                                                                                  \\ \cline{2-6} 
                                 & \textbf{JPEG (Q=50)} & \textbf{Cropout (30\%)} & \textbf{Dropout (30\%)} & \textbf{Crop (3.5\%)} & \textbf{Gaussian Filter($\sigma$=2)} \\ \hline
\textbf{HiDDeN \cite{Hidden}}         & 37.0                 & 6.0                     & 7.0                     & 12.0                  & 4.0                           \\ \hline
\textbf{\paperName{} (MT-NET, $\alpha\mathbf{=1}$)}         & 25.4                 & 7.5                     & 8.0                     & 0.0                   & 50.0                          \\ \hline
\end{tabular}
\end{adjustbox}
\label{tab:Hidden}
\end{table}

The other competitors are Random Matching Pursuit \cite{p36}. Since they have reported BER results for JPEG attack, we only present the relevant results for this attack. The comparisons are conducted with the two networks JT-Net and MT-Net against their best results on Granada dataset \cite{dataset} which is used by them to report their results. 1024 bit random watermarks are embedded into $512 \times 512$ dataset images, and BER values are reported. The strength factor of the networks is specifically set to adjust the SSIM to the same value of the competitor's. Then BER values of the extracted watermarks are compared in similar watermark qualities in terms of SSIM metric, while our PSNR is still better. Table \ref{tab:pursuit} shows that \paperName{} outperforms Random Matching Pursuit in terms of robustness (BER).
\begin{table}[!t]
\centering
\caption{\small Comparative results of robustness (BER) to Jpeg attack with three different qualities. In all networks the SSIM is nearly the same to that of \cite{p36} for a fair comparison.}
\begin{adjustbox}{width=0.6\columnwidth}
\begin{tabular}{cccc}
\hline
\multirow{2}{*}{\textbf{Method}}          & \multicolumn{3}{c}{\textbf{JPEG}}                                              \\ \cline{2-4} 
                                          & \textbf{50}              & \textbf{70}              & \textbf{90}              \\ \hline
\textbf{Random Matching Pursuit\cite{p36}} & 26.74                    & 17.64                    & 1.98                     \\ \hline
\textbf{\paperName{} (MT-NET), $\alpha=0.4$}                & 25.68                    & 16.49                    & 9.30                     \\ \hline
\textbf{\paperName{} (JT-NET),  $\alpha=0.8$}               & \multicolumn{1}{l}{6.75} & \multicolumn{1}{l}{2.32} & \multicolumn{1}{l}{1.31} \\ \hline
\end{tabular}
\end{adjustbox}
\label{tab:pursuit}
\end{table}
%##################################

\subsection{Diffusion pattern / data sharing}
\label{sec:expanalysis}

\begin{figure}[tb]
  \centering
    \includegraphics[width=1.0\columnwidth,height=5cm,keepaspectratio]{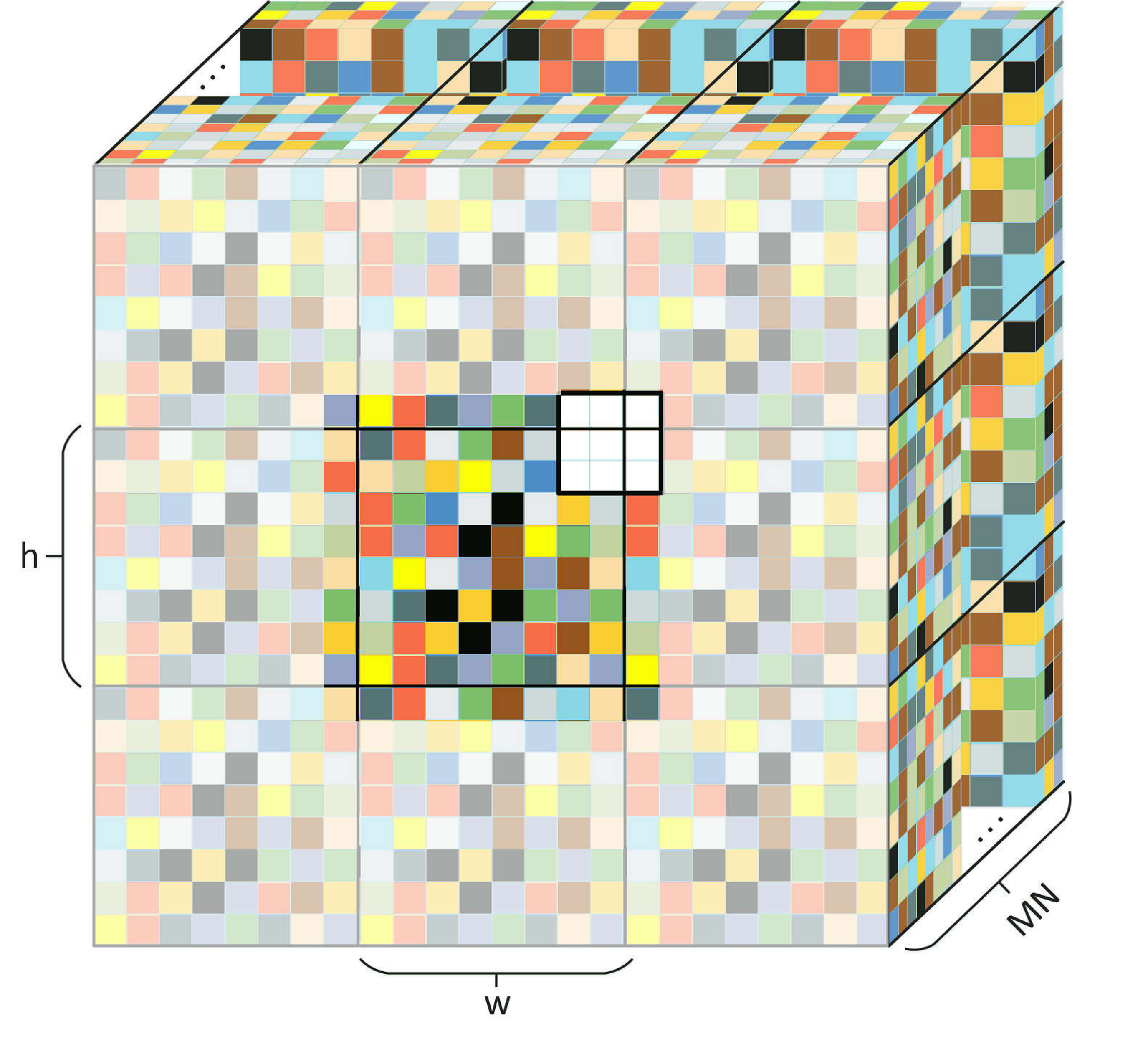}
    \caption{\small Circular convolution effect for the central tensor block ($MN\times h\times w$), is equivalent to repeating the tensor around itself and applying normal convolution. In this figure, the 8 faded tensors are repeated versions of the central tensor.}
	\label{fig:sharing}
\end{figure}

In this section, we demonstrate the ability of our networks in data sharing and diffusion among neighboring blocks. In our framework, we strategically use circular convolution to avoid zero-padding of the feature maps, to enhance the watermark strength and robustness. Fig. \ref{fig:sharing} demonstrates how the application of circular convolution leads to further diffusion and data sharing on the image borders. The circular convolution mask (white window) in Fig. \ref{fig:sharing} sweeps the opposite block edges, whenever it is on the borders. Consequently, all the neurons equally share the watermark data and the descending effect of zero-padding on watermarking will be avoided.
%##########################
\begin{figure*}[!t]
% First Row
\begin{subfigure}[!t]{1.0\textwidth}
\centering
%	\hspace{0.7cm}
	\begin{subfigure}[t]{.3\textwidth}
  		\caption{\scriptsize GT-NET}
 	 	\includegraphics[width=.8\linewidth]{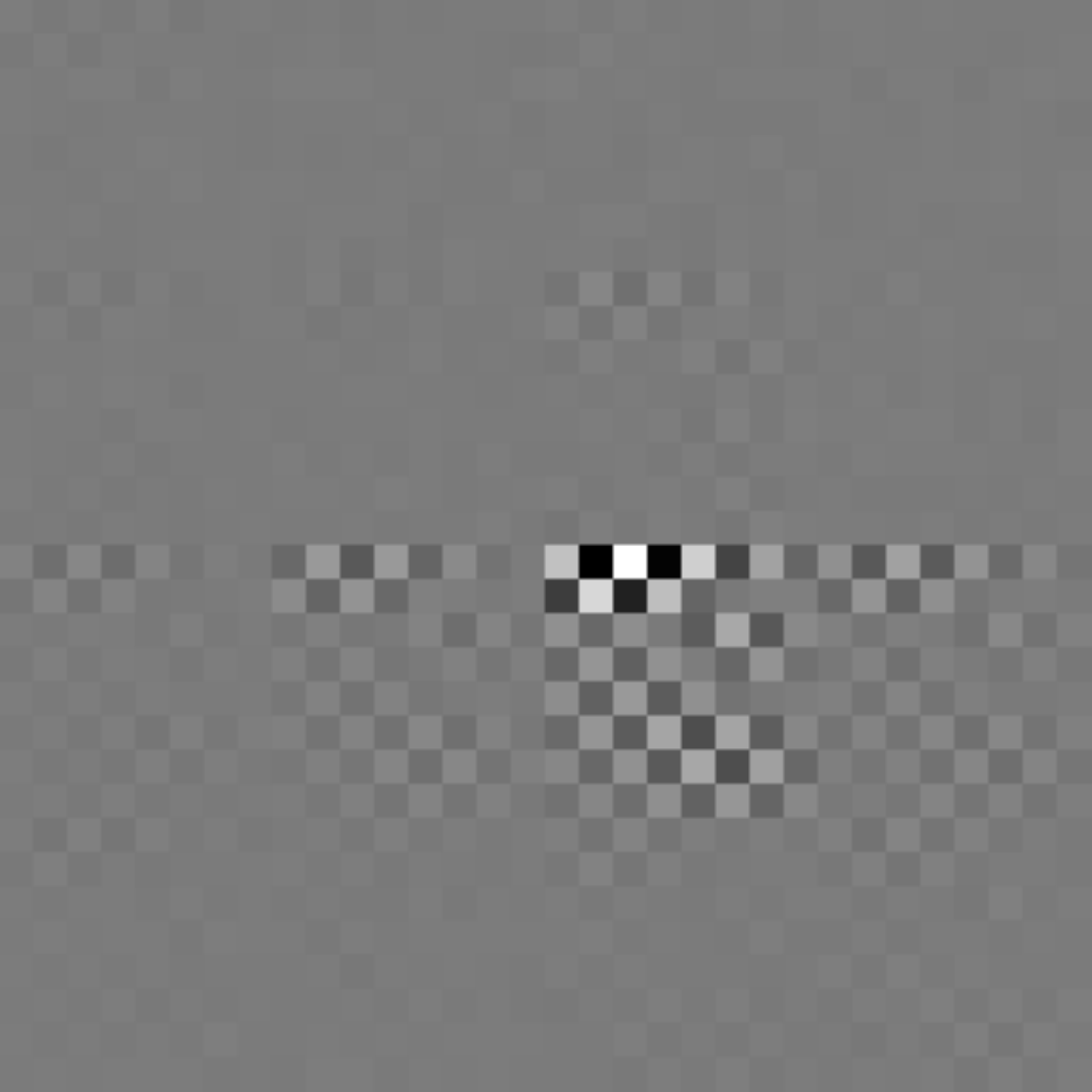}
  		\label{fig:sfig_patternGT}
	\end{subfigure}
%	\hspace{0.5cm}
\centering
	\begin{subfigure}[t]{.3\textwidth}
  		\caption{\scriptsize JT-NET}
 	 	\includegraphics[width=.8\linewidth]{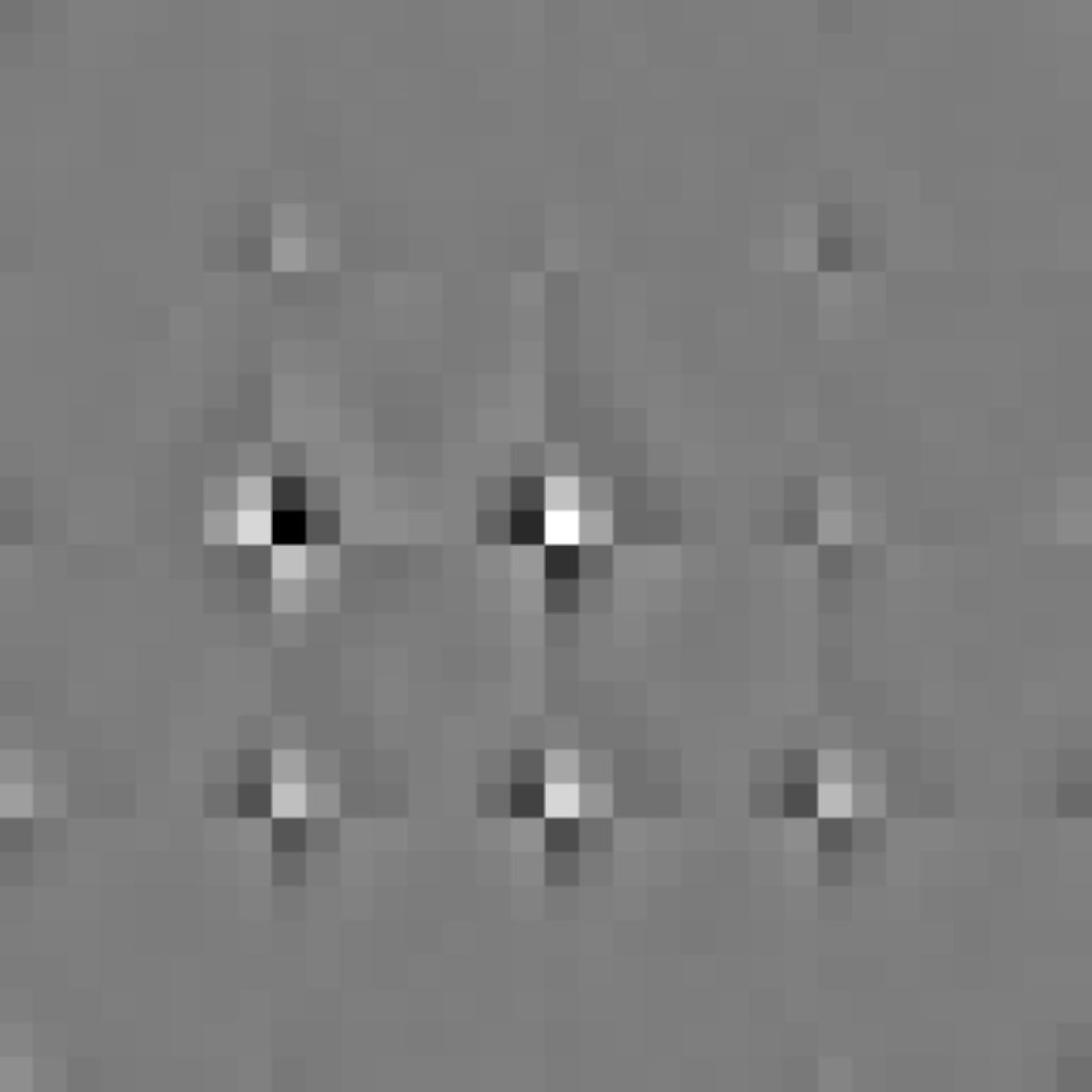}
  		\label{fig:sfig_patternJT}
	\end{subfigure}
%	\hspace{0.5cm}
\centering
	\begin{subfigure}[t]{.3\textwidth}
  		\caption{\scriptsize MT-NET}
 	 	\includegraphics[width=.8\linewidth]{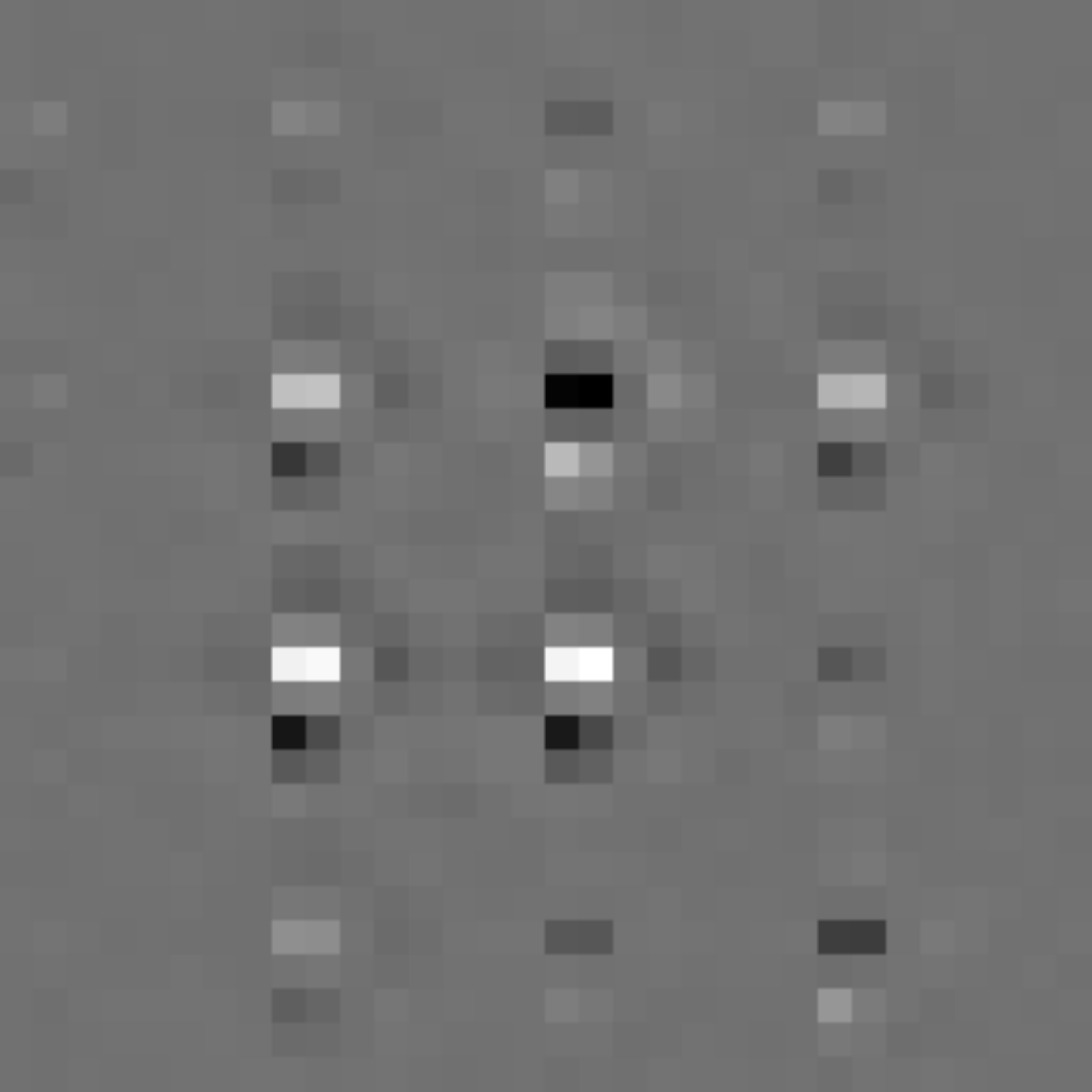}
  		\label{fig:sfig_patternMT}
	\end{subfigure}
\end{subfigure}
% Second Row
%\vspace{10pt}
\begin{subfigure}[ht]{1.0\textwidth}
	%\hspace{-0.5cm}
	\begin{subfigure}[t]{.3\textwidth}
 	 	\includegraphics[width=1.1\linewidth]{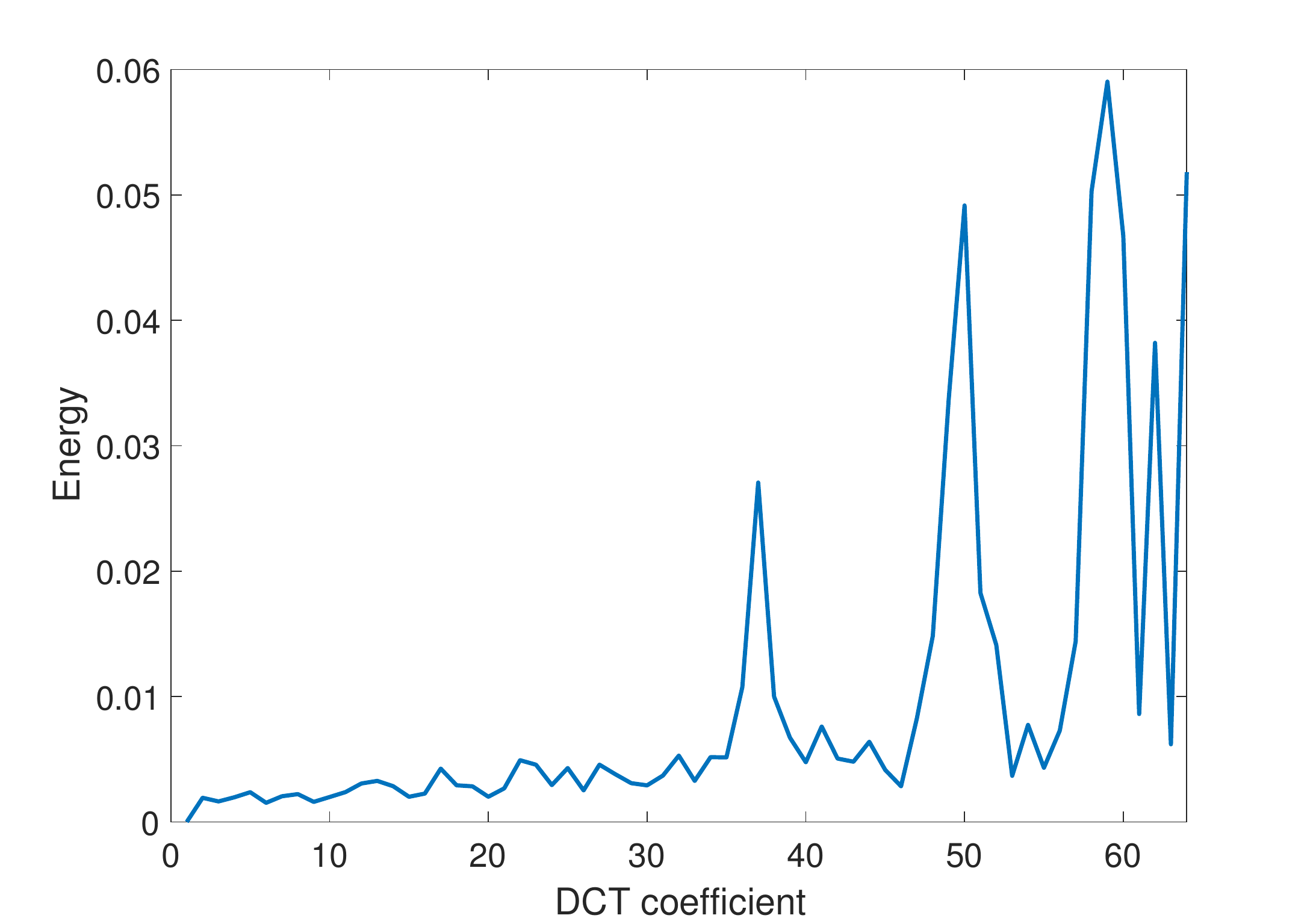}
  		\label{fig:sfig_specGT}
	\end{subfigure}
%	\hspace{0.35cm}
	\begin{subfigure}[t]{.3\textwidth}
 	 	\includegraphics[width=1.1\linewidth]{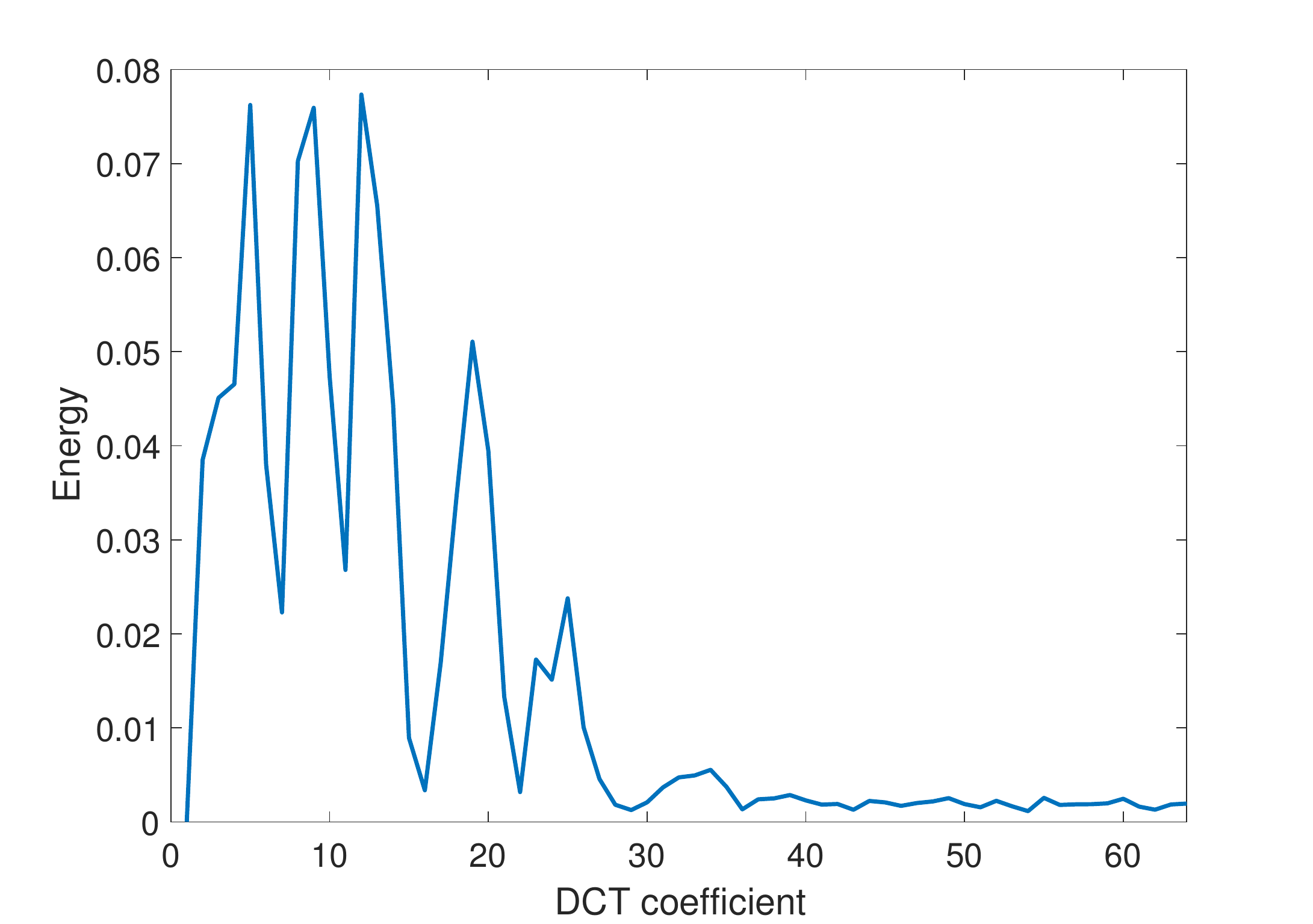}
  		\label{fig:sfig_specJT}
	\end{subfigure}
%	\hspace{0.4cm}
	\begin{subfigure}[t]{.3\textwidth}
 	 	\includegraphics[width=1.1\linewidth]{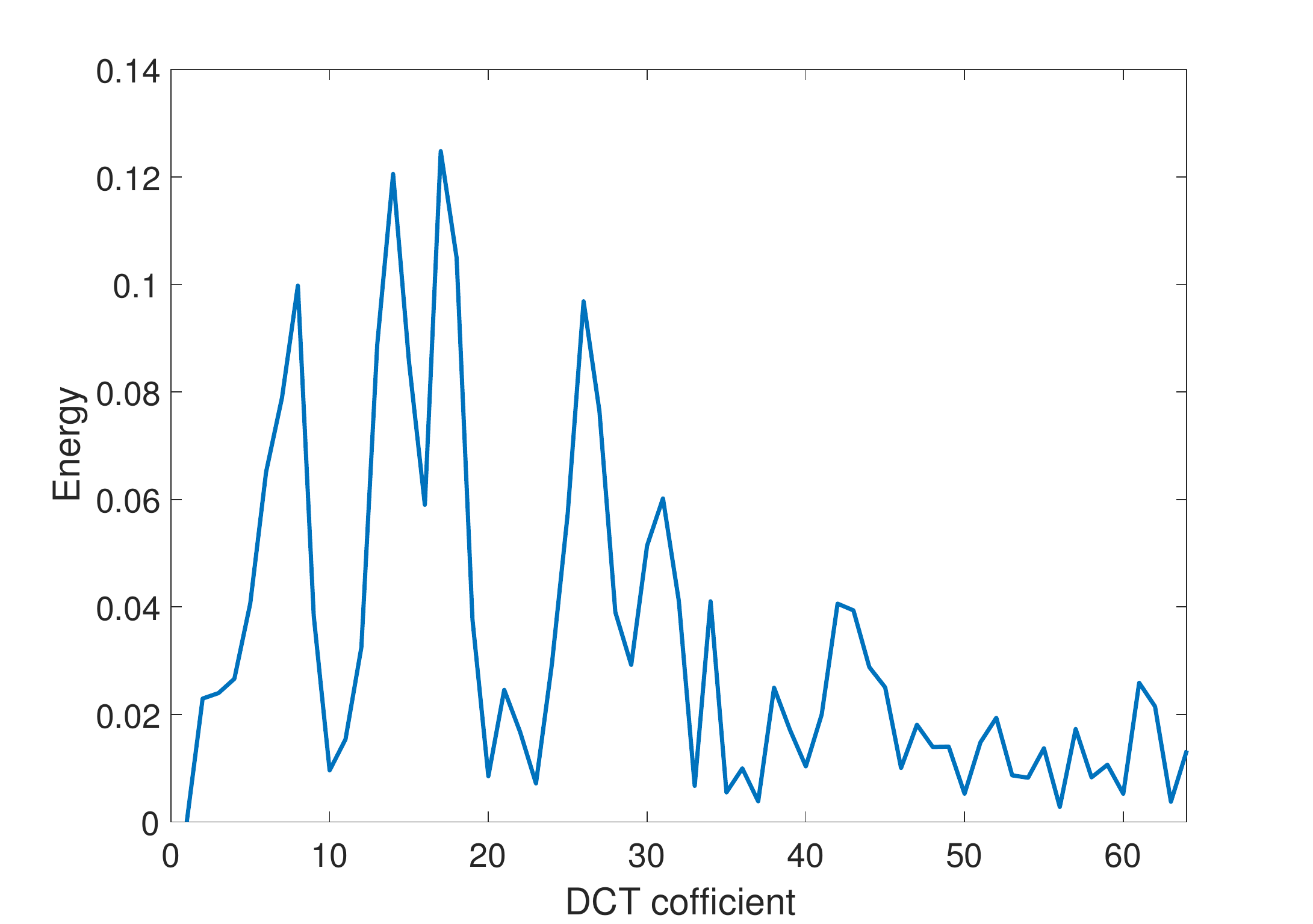}
  		\label{fig:sfig_specMT}
	\end{subfigure}
\end{subfigure}

\caption{\small First row: Watermark diffusion patterns of three networks ; Second row: Accumulated frequency energy curves of the networks.}
\label{fig:pattern}
\end{figure*}
To investigate the watermark patterns generated in neighboring blocks, we design a simple experiment. A watermark mask is formed with only one non-zero bit. Then we check the produced patterns by the embedding networks on a constant cover image, in which all the pixels are set to 128. However, this pattern alone does not give us enough insight without considering the effect of embedded zeroes. This is because the networks embed separate symbols for 0 and 1. Thus in the second step, we do the same experiment with another watermark mask of all zero bits. We call the difference of the produced watermarked images in the two mentioned experiments as \emph{diffusion pattern}. 
The produced patterns are invariant to the location of embedded non-zero bit, \ie, the same $8 \times 8$ block patterns are shifted according to the location of embedded 1.
The \emph{diffusion patterns} of the three trained networks are illustrated in Fig. \ref{fig:pattern}.

%\newpage
As illustrated in the \emph{diffusion patterns}, employing circular convolution leads to scattering the watermark data across the whole image, \ie, the $4\times 4$ watermark bits are diffused across the $32\times 32$ image block. The second row of Fig. \ref{fig:pattern}, demonstrates frequency energy curves of the \emph{diffusion patterns}, which are calculated by summing absolute values of DCT coefficients of all sixteen $8\times 8$ blocks of the \emph{diffusion pattern}. The accumulated DCT coefficients are arranged on the horizontal axis in zigzag order of the DCT block, \ie, starting from DC coefficient $DCT(0,0)$ towards the highest frequency $DCT(7,7)$. With a closer look to \emph{diffusion patterns} and their frequency energy curves in the DCT domain, it can be concluded that GT-NET embeds the watermark in high frequency coefficients of the cover image, JT-NET embeds in low frequency coefficients. However, MT-NET employs a more distributed embedding strategy with more concentration on low and middle frequency bands. 
It is worth to mention that the \emph{diffusion pattern} frequency curves are invariant to the location of embedded 1.
As demonstrated in Fig. \ref{fig:pattern}, all the networks have learnt to avoid embedding in DC coefficient, due to destructive effects on image quality and PSNR. 
\section{Conclusion}
\label{sec:Conclusion}
In this paper, we presented \paperName{}, an adaptive diffusion watermarking framework, composed of two Fully Convolutional Neural Networks with residual connections. Deep neural networks can be used to optimize existing algorithms. But in our work, the mentioned networks handle embedding and extraction processes. In other words, these networks learn a new watermarking algorithm in any desired transform domain. The networks were trained end-to-end to conduct a blind secure watermarking in the desired domain. The framework can be customized for the level of robustness vs. imperceptibility by tunable parameters during training and test of the networks. The proposed framework simulates various attacks as a differentiable network layer to facilitate end-to-end training. For instance, a differentiable approximation of JPEG attack is developed, which largely improves the watermarking robustness to this attack. In this work, we presented three network instances of \paperName{}, each of which trained under different attacks. We demonstrated the different nature/behavior of the trained networks, where each of them embeds in particular spectral regions due to their different training strategies. An important characteristic of the suggested system, which leads to improved security and robustness, is its capability to diffuse/share watermark data among a relatively wide area of the cover image. We visually illustrated this data sharing and diffusion-watermarking using diffusion patterns. Furthermore, we proposed two attacks to experimentally prove the effect of this unique characteristic on improving the watermarking robustness. Comparative results against recent state-of-the-art works demonstrate the superiority of \paperName{} in terms of imperceptibility and robustness.



\vspace{40pt}
\bibliographystyle{elsarticle-num}
\bibliography{refs}

\begin{thebibliography}{10}
\expandafter\ifx\csname url\endcsname\relax
  \def\url#1{\texttt{#1}}\fi
\expandafter\ifx\csname urlprefix\endcsname\relax\def\urlprefix{URL }\fi
\expandafter\ifx\csname href\endcsname\relax
  \def\href#1#2{#2} \def\path#1{#1}\fi

\bibitem{p1}
W.~Szepanski, {A signal theoretic method for creating forgery-proof documents
  for automatic verification} 101~(109) (1979) 368.

\bibitem{p2}
I.~Cox, M.~Miller, J.~Bloom, J.~Fridrich, T.~Kalker, {Digital watermarking and
  steganography}, Morgan kaufmann, 2007.

\bibitem{p3}
A.~Shehab, M.~Elhoseny, K.~Muhammad, A.~K. Sangaiah, P.~Yang, H.~Huang, G.~Hou,
  {Secure and robust fragile watermarking scheme for medical images}, IEEE
  Access 6 (2018) 10269--10278.

\bibitem{p4}
Y.~Cheng, {Music database retrieval based on spectral similarity}, 2nd Int.
  Sympo-sium on Music Information Retrieval (IS-MIR), Oct., 2001.

\bibitem{p5}
M.~Faundez-Zanuy, M.~Hagm{\"u}ller, G.~Kubin, Speaker identification security
  improvement by means of speech watermarking, Pattern recognition 40~(11)
  (2007) 3027--3034.

\bibitem{p6}
R.~S. Broughton, W.~C. Laumeister, {Interactive video method and apparatus}
  (1989).

\bibitem{p7}
M.~Hagm{\"{u}}ller, H.~Hering, A.~Kr{\"{o}}pfl, G.~Kubin, {Speech watermarking
  for air traffic control}, Watermark 8~(9) (2004) 10.

\bibitem{p9}
F.~N. Thakkar, V.~K. Srivastava, {A blind medical image watermarking: DWT-SVD
  based robust and secure approach for telemedicine applications}, Multimedia
  Tools and Applications 76~(3) (2017) 3669--3697.

\bibitem{p10}
O.~Benrhouma, H.~Hermassi, A.~A.~A. El-Latif, S.~Belghith, {Chaotic watermark
  for blind forgery detection in images}, Multimedia Tools and Applications
  75~(14) (2016) 8695--8718.

\bibitem{p11}
D.~G. Savakar, A.~Ghuli,
  \href{http://link.springer.com/10.1134/S1054661817030257}{{Non-blind digital
  watermarking with enhanced image embedding capacity using DMeyer wavelet
  decomposition, SVD, and DFT}}, Pattern Recognition and Image Analysis 27~(3)
  (2017) 511--517.
\newblock \href {https://doi.org/10.1134/S1054661817030257}
  {\path{doi:10.1134/S1054661817030257}}.
\newline\urlprefix\url{http://link.springer.com/10.1134/S1054661817030257}

\bibitem{p12}
G.~Anbarjafari, C.~Ozcinar, {Imperceptible non-blind watermarking and
  robustness against tone mapping operation attacks for high dynamic range
  images}, Multimedia Tools and Applications 77~(18) (2018) 24521--24535.
\newblock \href {https://doi.org/10.1007/s11042-018-5759-1}
  {\path{doi:10.1007/s11042-018-5759-1}}.

\bibitem{p14}
A.~M. Abdelhakim, M.~Abdelhakim, {A time-efficient optimization for robust
  image watermarking using machine learning}, Expert Systems with Applications
  100 (2018) 197--210.

\bibitem{p15}
Z.~Zhi-Ming, L.~Rong-Yan, W.~Lei, {Adaptive watermark scheme with RBF neural
  networks}, in: Neural Networks and Signal Processing, 2003. Proceedings of
  the 2003 International Conference on, Vol.~2, IEEE, 2003, pp. 1517--1520.

\bibitem{p16}
L.~Sanping, Z.~Yusen, Z.~Hui, {A wavelet-domain watermarking technique based on
  support vector regression}, in: Grey Systems and Intelligent Services, 2007.
  GSIS 2007. IEEE International Conference on, IEEE, 2007, pp. 1112--1116.

\bibitem{p17}
A.~Khan, S.~F. Tahir, A.~Majid, T.-S. Choi, Machine learning based adaptive
  watermark decoding in view of anticipated attack, Pattern Recognition 41~(8)
  (2008) 2594--2610.

\bibitem{p18}
R.~P. Singh, N.~Dabas, V.~Chaudhary, {Online sequential extreme learning
  machine for watermarking in DWT domain}, Neurocomputing 174 (2016) 238--249.

\bibitem{Heidari}
M.~Heidari, S.~Samavi, S.~M.~R. Soroushmehr, S.~Shirani, N.~Karimi,
  K.~Najarian, {Framework for robust blind image watermarking based on
  classification of attacks}, Multimedia Tools and Applications 76~(22) (2017)
  23459--23479.

\bibitem{SVM}
E.~Pasolli, F.~Melgani, D.~Tuia, F.~Pacifici, W.~J. Emery, {SVM active learning
  approach for image classification using spatial information}, IEEE
  Transactions on Geoscience and Remote Sensing 52~(4) (2014) 2217--2233.

\bibitem{SVR}
M.~Narwaria, W.~Lin,
  \href{http://www.ncbi.nlm.nih.gov/pubmed/20100674}{{Objective image quality
  assessment based on support vector regression.}}, IEEE transactions on neural
  networks / a publication of the IEEE Neural Networks Council 21~(3) (2010)
  515--9.
\newblock \href {https://doi.org/10.1109/TNN.2010.2040192}
  {\path{doi:10.1109/TNN.2010.2040192}}.
\newline\urlprefix\url{http://www.ncbi.nlm.nih.gov/pubmed/20100674}

\bibitem{RBFNN}
W.~W.~Y. Ng, A.~Dorado, D.~S. Yeung, W.~Pedrycz, E.~Izquierdo, {Image
  classification with the use of radial basis function neural networks and the
  minimization of the localized generalization error}, Pattern Recognition
  40~(1) (2007) 19--32.

\bibitem{KNN}
L.~Ma, M.~M. Crawford, J.~Tian, {Local manifold learning-based $ k
  $-nearest-neighbor for hyperspectral image classification}, IEEE Transactions
  on Geoscience and Remote Sensing 48~(11) (2010) 4099--4109.

\bibitem{p20}
Z.~Zheng, L.~Zheng, Y.~Yang, {A discriminatively learned CNN embedding for
  person reidentification}, ACM Transactions on Multimedia Computing,
  Communications, and Applications (TOMM) 14~(1) (2017) 13.

\bibitem{p21}
G.~Huang, Z.~Liu, L.~{Van Der Maaten}, K.~Q. Weinberger, {Densely Connected
  Convolutional Networks.} 1~(2) (2017) 3.

\bibitem{p22}
S.~Ren, K.~He, R.~Girshick, J.~Sun, {Faster R-CNN: Towards real-time object
  detection with region proposal networks}, in: Advances in neural information
  processing systems, 2015, pp. 91--99.

\bibitem{p24}
H.~Kandi, D.~Mishra, S.~R.~S. Gorthi, Exploring the learning capabilities of
  convolutional neural networks for robust image watermarking, Computers \&
  Security 65 (2017) 247--268.

\bibitem{p25}
S.-M. Mun, S.-H. Nam, H.-U. Jang, D.~Kim, H.-K. Lee, {A robust blind
  watermarking using convolutional neural network}, arXiv preprint
  arXiv:1704.03248.

\bibitem{Hidden}
J.~Zhu, R.~Kaplan, J.~Johnson, L.~Fei-Fei, {HiDDeN: Hiding data with deep
  networks}, arXiv preprint arXiv:1807.09937.

\bibitem{GAN}
I.~Goodfellow, J.~Pouget-Abadie, M.~Mirza, B.~Xu, D.~Warde-Farley, S.~Ozair,
  A.~Courville, Y.~Bengio, {Generative adversarial nets}, in: Advances in
  neural information processing systems, 2014, pp. 2672--2680.

\bibitem{Resnet}
K.~He, X.~Zhang, S.~Ren, J.~Sun, {Deep residual learning for image
  recognition}, in: Proceedings of the IEEE conference on computer vision and
  pattern recognition, 2016, pp. 770--778.

\bibitem{p28}
R.-Z. Wang, C.-F. Lin, J.-C. Lin, {Image hiding by optimal LSB substitution and
  genetic algorithm}, Pattern recognition 34~(3) (2001) 671--683.

\bibitem{p29}
S.~A. Parah, J.~A. Sheikh, N.~A. Loan, G.~M. Bhat, {Robust and blind
  watermarking technique in DCT domain using inter-block coefficient
  differencing}, Digital Signal Processing 53 (2016) 11--24.

\bibitem{p30}
A.~K. Singh, M.~Dave, A.~Mohan, {Robust and secure multiple watermarking in
  wavelet domain}, Journal of medical imaging and health informatics 5~(2)
  (2015) 406--414.

\bibitem{p31}
E.~Etemad, S.~Samavi, S.~R. Soroushmehr, N.~Karimi, M.~Etemad, S.~Shirani,
  K.~Najarian, Robust image watermarking scheme using bit-plane of hadamard
  coefficients, Multimedia Tools and Applications 77~(2) (2018) 2033--2055.

\bibitem{p32}
S.~Etemad, M.~Amirmazlaghani, A new multiplicative watermark detector in the
  contourlet domain using t location-scale distribution, Pattern Recognition 77
  (2018) 99--112.

\bibitem{Fazlali}
H.~Fazlali, S.~Samavi, N.~Karimi, S.~Shirani, {Adaptive blind image
  watermarking using edge pixel concentration}, Multimedia Tools and
  Applications 76~(2) (2017) 3105--3120.

\bibitem{p34}
H.~Sadreazami, M.~O. Ahmad, M.~N.~S. Swamy, {Multiplicative watermark decoder
  in contourlet domain using the normal inverse Gaussian distribution}, IEEE
  Transactions on Multimedia 18~(2) (2016) 196--207.

\bibitem{p35}
N.~M. Makbol, B.~E. Khoo, T.~H. Rassem, K.~Loukhaoukha, {A new reliable
  optimized image watermarking scheme based on the integer wavelet transform
  and singular value decomposition for copyright protection}, Information
  Sciences 417 (2017) 381--400.

\bibitem{p36}
G.~Hua, L.~Zhao, H.~Zhang, G.~Bi, Y.~Xiang, {Random matching pursuit for image
  watermarking}, IEEE Transactions on Circuits and Systems for Video
  Technology.

\bibitem{p37}
J.~Li, C.~Yu, B.~B. Gupta, X.~Ren, {Color image watermarking scheme based on
  quaternion Hadamard transform and Schur decomposition}, Multimedia Tools and
  Applications 77~(4) (2018) 4545--4561.

\bibitem{p38}
X.~Liu, G.~Han, J.~Wu, Z.~Shao, G.~Coatrieux, H.~Shu, {Fractional Krawtchouk
  transform with an application to image watermarking}, IEEE Transactions on
  Signal Processing 65~(7) (2017) 1894--1908.

\bibitem{p39}
B.~Chen, G.~W. Wornell, {Quantization index modulation: A class of provably
  good methods for digital watermarking and information embedding}, IEEE
  Transactions on Information Theory 47~(4) (2001) 1423--1443.

\bibitem{p40}
I.~Daubechies, W.~Sweldens, {Factoring wavelet transforms into lifting steps},
  Journal of Fourier analysis and applications 4~(3) (1998) 247--269.

\bibitem{p41}
P.~Dabas, K.~Khanna, {A study on spatial and transform domain watermarking
  techniques}, International journal of computer applications 71~(14).

\bibitem{Dropout}
N.~Srivastava, G.~Hinton, A.~Krizhevsky, I.~Sutskever, R.~Salakhutdinov,
  {Dropout: a simple way to prevent neural networks from overfitting}, The
  Journal of Machine Learning Research 15~(1) (2014) 1929--1958.

\bibitem{Tensorflow}
M.~Abadi, P.~Barham, J.~Chen, Z.~Chen, A.~Davis, J.~Dean, M.~Devin,
  S.~Ghemawat, G.~Irving, M.~Isard, {Tensorflow: a system for large-scale
  machine learning.}, in: OSDI, Vol.~16, 2016, pp. 265--283.

\bibitem{CIFAR}
A.~Krizhevsky, V.~Nair, G.~Hinton, {The CIFAR-10 dataset}, online: http://www.
  cs. toronto. edu/kriz/cifar. html.

\bibitem{Pascal}
M.~Everingham, L.~Van{\~{}}Gool, C.~K.~I. Williams, J.~Winn, A.~Zisserman, {The
  PASCAL visual object classes challenge 2012 (VOC2012) results},
  http://www.pascal-network.org/challenges/VOC/voc2012/workshop/index.html.

\bibitem{dataset}
\href{http://decsai.ugr.es/cvg/CG/base.htm}{{Dataset of standard 512$\times$512
  grayscale test images}}.
\newline\urlprefix\url{http://decsai.ugr.es/cvg/CG/base.htm}

\bibitem{COCO}
T.-Y. Lin, M.~Maire, S.~Belongie, J.~Hays, P.~Perona, D.~Ramanan,
  P.~Doll{\'{a}}r, C.~L. Zitnick, {Microsoft coco: Common objects in context},
  in: European conference on computer vision, Springer, 2014, pp. 740--755.

\bibitem{ELU}
D.-A. Clevert, T.~Unterthiner, S.~Hochreiter,
  \href{http://arxiv.org/abs/1511.07289}{{Fast and accurate deep network
  learning by exponential linear units (ELUs)}}, arXiv preprint
  arXiv:1511.07289\href {http://arxiv.org/abs/1511.07289}
  {\path{arXiv:1511.07289}}, \href
  {https://doi.org/10.3233/978-1-61499-672-9-1760}
  {\path{doi:10.3233/978-1-61499-672-9-1760}}.
\newline\urlprefix\url{http://arxiv.org/abs/1511.07289}

\bibitem{p49}
Y.~Xue, X.~Liao, L.~Carin, B.~Krishnapuram, Multi-task learning for
  classification with dirichlet process priors, Journal of Machine Learning
  Research 8~(Jan) (2007) 35--63.

\end{thebibliography}
\newpage
\appendix%[Proofs of The Lemmas]
\section{}
\label{sec:lemmas}
\subsection{Expansion of equation \eqref{prelema} for DCT transform:}
If $\mathbf{f_T}_{M\times N}$ is the two dimensional DCT transform of $\mathbf{f}_{M\times N}$, it can be written as:
\begin{align}
\label{lemma11}
\mathbf{f_T}(u,v)=\frac{1}{MN}\sum_{m=0}^{M-1}\sum_{n=0}^{N-1}{\mathbf{f}(m,n)}\cos{\left( \frac{\left(2m+1\right) u\pi}{2M} \right)}\nonumber\\
\cos{\left( \frac{\left(2n+1\right) v\pi}{2N} \right)} 
\end{align}
By reshaping the matrices $\mathbf{f}$ and $\mathbf{f_T}$ to vectors of length $MN$, the $\theta^{th}$ element of the vector  $\mathbf{f_T}$ can be written by change of variables of $m=\left\lfloor  \frac{k}{M} \right\rfloor$, $n=k-mM$, $u=\left\lfloor\frac{\theta}{M}\right\rfloor$ and $v=\theta-uM$.
\begin{align}
\label{lemma12}
\mathbf{f_T}(\theta)=\frac{1}{MN}\sum_{k=0}^{MN-1}\mathbf{f}(k)  &\cos{\left(\frac{\left( 2\left\lfloor\frac{k}{M}\right\rfloor+1\right)\left\lfloor\frac{\theta}{M}\right\rfloor\pi}{2M}\right)}\nonumber\\
&\cos{\left( \frac{\left(2\left( k- \left\lfloor  \frac{k}{M} \right\rfloor M  \right)+1\right) \left( \theta-\left\lfloor\frac{\theta}{M} \right\rfloor M\right)\pi}{2N} \right)}   
\end{align}
in which $\mathbf{f}(k)$ is the $k^{th}$ element of the vector $\mathbf{f}$ and so:
%%%%%%%%%%
\begin{equation} 
\label{lemma13}
\mathbf{f_T}(\theta)=\sum_{k=0}^{MN-1}{ \mathbf{f}(k)\mathbf{D}(\theta,k) }
\end{equation}
%or
%\begin{equation} 
%\label{lemma14}
% \mathbf{f_T}_{MN\times 1} =\mathbf{D}_{MN \times MN} \mathbf{f}_{MN \times 1}
%\end{equation}
in which:
\begin{eqnarray}
\label{lemma15}
\mathbf{D}(\theta ,k) = \frac{1}{MN} \cos{\left(\frac{\left( 2\left\lfloor\frac{k}{M}\right\rfloor+1\right)\left\lfloor\frac{\theta}{M}\right\rfloor\pi}{2M}\right)}\nonumber\\
\cos{\left( \frac{\left(2\left( k- \left\lfloor  \frac{k}{M} \right\rfloor M  \right)+1\right) \left( \theta-\left\lfloor\frac{\theta}{M} \right\rfloor M\right)\pi}{2N} \right)}
\end{eqnarray}
%%
%%%%%%%%%%%%%%%%%%%%

%%
\subsection{Expansion of equation \eqref{prelema} for Hadamard transform:}
Hadamard transform of an $N\times N$ block $\mathbf{f}$ is defined by $\mathbf{f_T}= \mathbf{HfH}$, where $\mathbf{H}$ is the $N\times N$ Hadamard matrix. 
Elements of the transformed matrix $\mathbf{f_T}$ are calculated by:
%\begin{equation} \label{lemma21}
% \mathbf{f_T}= \mathbf{HfH}
%\end{equation}
%That $\mathbf{H}$ is $N\times N$ Hadamard matrix. So:
\begin{equation} \label{lemma22}
 \mathbf{f_T}(u,v)= \sum_{n=0}^{N-1}{\mathbf{H}(u,n)  \mathbf{A}(n,v)}
\end{equation}
where
\begin{equation} \label{lemma23}
 \mathbf{A}(n,v)= \sum_{m=0}^{N-1}{\mathbf{f}(n,m)  \mathbf{H}(m,v)}
\end{equation}
%%%%%%%%%%%
\begin{eqnarray}
\label{lemma24}
 \mathbf{f_T}(u,v)= \sum_{n=0}^{N-1}{\mathbf{H}(u,n)  \sum_{m=0}^{N-1}{\mathbf{f}(n,m)  \mathbf{H}(m,v)}} =\nonumber\\  \sum_{m=0}^{N-1}  \sum_{n=0}^{N-1}   \mathbf{f}(n,m) \mathbf{H}(u,n) \mathbf{H}(m,v)
\end{eqnarray}
Equation \eqref{lemma24} is reshaped by change of variables $m=\left\lfloor  \frac{k}{N} \right\rfloor$, $n=k-mN$, $u=\left\lfloor\frac{\theta}{N}\right\rfloor$ and $v=\theta-uN$, as bellow:
\begin{equation} \label{lemma25}
\mathbf{f_T}(\theta)=\sum_{k=0}^{N^2-1}{ \mathbf{f}(k)\mathbf{D}(\theta,k) }
\end{equation}
%or
%\begin{equation} \label{lemma26}
% \mathbf{f_T}_{N^2\times 1} =\mathbf{D}_{N^2 \times N^2} \mathbf{f}_{N^2 \times 1}
%\end{equation}
where:
\begin{align}
 \label{lemma27}
&\mathbf{D}(\theta ,k) =\mathbf{H}(u,n) \mathbf{H}(m,v)=\nonumber\\
&\mathbf{H}\left( \left\lfloor\frac{\theta}{N} \right\rfloor,k-\left\lfloor\frac{k}{N} \right\rfloor N\right)    \mathbf{H}\left( \left\lfloor\frac{k}{N} \right\rfloor,\theta-\left\lfloor\frac{\theta}{N} \right\rfloor N  \right)
\end{align}



%%\bibitem{}

%%\end{thebibliography}
\end{document}